\documentclass[sigconf]{acmart}
\AtBeginDocument{%
  }

\copyrightyear{2026}
\acmYear{2026}
\setcopyright{cc}
\setcctype{by}
\acmConference[CHI '26]{Proceedings of the 2026 CHI Conference on Human Factors in Computing Systems}{April 13--17, 2026}{Barcelona, Spain}
\acmBooktitle{Proceedings of the 2026 CHI Conference on Human Factors in Computing Systems (CHI '26), April 13--17, 2026, Barcelona, Spain}
\acmPrice{}
\acmDOI{10.1145/3772318.3790953}
\acmISBN{979-8-4007-2278-3/2026/04}

\usepackage{enumitem}
\usepackage{tabularx}
\usepackage{booktabs}
\usepackage{tikz}
\usetikzlibrary{positioning,shapes,arrows}

\PassOptionsToPackage{svgnames,x11names,dvipsnames}{xcolor}

\usepackage{xcolor}

\definecolor{casedesc}{HTML}{5EE9B5}
\definecolor{riskpred}{HTML}{8EC5FF}
\definecolor{peeract}{HTML}{FFB86A}
\definecolor{plans}{HTML}{FFA2A2}

\newcommand{\inlinebox}[2]{%
  \begingroup
  \setlength{\fboxsep}{2pt}%
  \fcolorbox{#1}{#1!50}{\textbf{#2}}%
  \endgroup
}

\makeatletter
\newcommand{\DeclareRule}[4]{%
  \expandafter\newcommand\csname #1\endcsname{%
    \@ifstar
      {\csname #1@star\endcsname}
      {\csname #1@nostar\endcsname}%
  }%
  \expandafter\newcommand\csname #1@star\endcsname[1]{#3}%
  \expandafter\newcommand\csname #1@nostar\endcsname[1]{#4}%
}
\makeatother

\DeclareRule
  {consistentfeatures}
  {}
  {\textbf{Consistent features of case}}
  {\inlinebox{casedesc}{R1}}

\DeclareRule
  {unusualfeatures}
  {}
  {\textbf{Unusual features of case}}
  {\inlinebox{casedesc}{R2}}

\DeclareRule
  {anyrisk}
  {}
  {\textbf{Risk score}}
  {\inlinebox{riskpred}{R3}}
\DeclareRule
  {lowhighrisk}
  {}
  {\textbf{Risk score: Low/High}}
  {\inlinebox{riskpred}{\textbf{R3-L/H}}}

\DeclareRule
  {moderaterisk}
  {}
  {\textbf{Risk score: Moderate}}
  {\inlinebox{riskpred}{\textbf{R3-M}}}

\DeclareRule
  {diffrisk}
  {}
  {\textbf{Difference in plan-dependent risk}}
  {\inlinebox{riskpred}{R4}}

\DeclareRule
  {diffriskyes}
  {}
  {\textbf{Difference in plan-dependent risk: yes}}
  {\inlinebox{riskpred}{R4-Y}}

\DeclareRule
  {diffriskno}
  {}
  {\textbf{Difference in plan-dependent risk: no}}
  {\inlinebox{riskpred}{\textbf{R4-N}}}

\DeclareRule
  {commonaction}
  {}
  {\textbf{Action common among peers}}
  {\inlinebox{peeract}{R5}}

\DeclareRule
  {commonactionyes}
  {}
  {\textbf{Action common among peers: yes}}
  {\inlinebox{peeract}{\textbf{R5-Y}}}

\DeclareRule
  {commonactionno}
  {}
  {\textbf{Action common among peers: no}}
  {\inlinebox{peeract}{\textbf{R5-N}}}

\DeclareRule
  {consensusaction}
  {}
  {\textbf{One consensus peer action}}
  {\inlinebox{peeract}{\textbf{R6}}}

\DeclareRule
  {consensusactionyes}
  {}
  {\textbf{One consensus peer action: yes}}
  {\inlinebox{peeract}{\textbf{R6-Y}}}

\DeclareRule
  {consensusactionno}
  {}
  {\textbf{One consensus peer action: no}}
  {\inlinebox{peeract}{\textbf{R6-N}}}

\DeclareRule
  {planmention}
  {}
  {\textbf{Plan mention}}
  {\inlinebox{plans}{\textbf{R7}}}
  
\DeclareRule
  {recommendedplan}
  {}
  {\textbf{Recommended plan}}
  {\inlinebox{plans}{R8}}

\begin{document}

\title[Intelligent Reasoning Cues]{Intelligent Reasoning Cues: A Framework and Case Study of the Roles of AI Information in Complex Decisions}

\author{Venkatesh Sivaraman}
\email{venkats@cmu.edu}
\orcid{0000-0002-6965-3961}
\affiliation{%
  \institution{Carnegie Mellon University}
  \city{Pittsburgh}
  \state{PA}
  \country{USA}
}

\author{Eric Mason}
\affiliation{%
  \institution{Carnegie Mellon University}
  \city{Pittsburgh}
  \state{PA}
  \country{USA}
}

\author{Mengfan Li}
\affiliation{%
  \institution{Carnegie Mellon University}
  \city{Pittsburgh}
  \state{PA}
  \country{USA}
}

\author{Jessica Tong}
\authornote{Work done while at Carnegie Mellon University.}
\affiliation{%
  \institution{Pomona College}
  \city{Claremont}
  \state{CA}
  \country{USA}
}

\author{Andrew J. King}
\affiliation{%
  \institution{University of Pittsburgh}
  \city{Pittsburgh}
  \state{PA}
  \country{USA}
}

\author{Jeremy M. Kahn}
\affiliation{%
  \institution{University of Pittsburgh}
  \city{Pittsburgh}
  \state{PA}
  \country{USA}
}

\author{Adam Perer}
\affiliation{%
  \institution{Carnegie Mellon University}
  \city{Pittsburgh}
  \state{PA}
  \country{USA}
}
\renewcommand{\shortauthors}{Sivaraman et al.}

\begin{abstract}
Artificial intelligence (AI)-based decision support systems can be highly accurate yet still fail to support users or improve decisions.
Existing theories of AI-assisted decision-making focus on calibrating reliance on AI advice, leaving it unclear how different system designs might influence the reasoning processes underneath.
We address this gap by reconsidering AI interfaces as collections of \textit{intelligent reasoning cues}: discrete pieces of AI information that can individually influence decision-making.
We then explore the roles of eight types of reasoning cues in a high-stakes clinical decision (treating patients with sepsis in intensive care).
Through contextual inquiries with six teams and a think-aloud study with 25 physicians, we find that reasoning cues have distinct patterns of influence that can directly inform design.
Our results also suggest that reasoning cues should prioritize tasks with high variability and discretion, adapt to ensure compatibility with evolving decision needs, and provide complementary, rigorous insights on complex cases.
\end{abstract}

\ccsdesc[500]{Human-centered computing~Interactive systems and tools}
\ccsdesc[500]{Applied computing~Health care information systems}
\ccsdesc[500]{Computing methodologies~Artificial intelligence}

\keywords{Clinical Reasoning, AI-Assisted Decision-Making, Decision Support, Explainable AI}
\begin{teaserfigure}
\includegraphics[width=\linewidth]{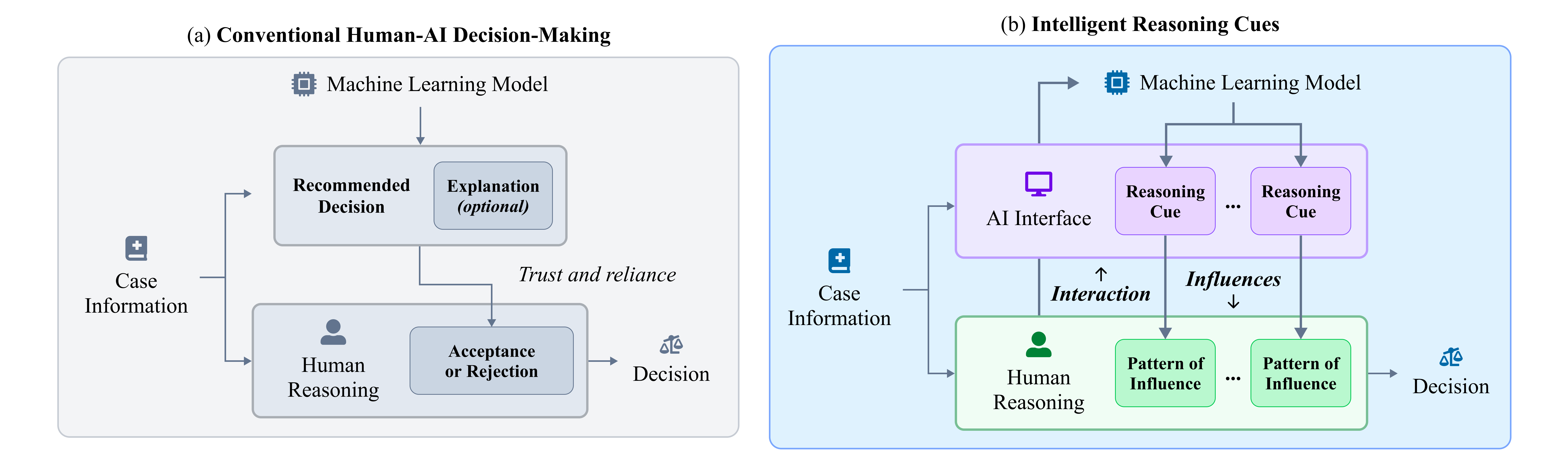}
\caption{(a) The conventional model used in most human-AI decision-making literature (e.g. \cite{vaccaro_when_2024,schemmer_appropriate_2023,Bansal2020}) is that a machine learning model produces a decision recommendation as well as an optional explanation, which the user either accepts or ignores in favor of their own judgment. (b) The intelligent reasoning cues framework generalizes AI interfaces as any collection of one or more reasoning cues, which can span many types of AI-derived information. These cues shape a human decision-maker's thinking by inducing specific patterns of influence on reasoning beyond simple reliance. Users' interactions with the interface can also drive the model to generate new reasoning cues that further support decision-making.}
\label{fig:cognitive-models}
\end{teaserfigure}


\maketitle

\section{Introduction}

Artificial intelligence (AI) tools promise to transform expert decision-making by improving the quality of decisions and saving time and effort.
While a number of AI-powered decision aids have seen success in deployment~\cite{adams_prospective_2022,Sendak2020} and decision-makers are increasingly open to AI support~\cite{newton_systematic_2025}, many tools still fail to effectively support the experts tasked with using them~\cite{greenes_clinical_2018}.
Human-centered AI research has an important role in identifying and mitigating these failures~\cite{sivaraman_tempo_2025,saxena_ai_2025}, ideally deriving generalizable principles that drive the effective design and evaluation of AI-based decision support.

The majority of AI-based decision support systems so far have focused on automating parts of a decision task, saving decision-makers time when they can rely on the system but causing additional burden otherwise. 
For example, predictive models in emergency medicine that forecast the onset of a disease or a patient's deterioration can improve the quality of health care by flagging cases as high-risk or in need of attention~\cite{hyland_early_2020}. 
However, the decision-makers in these contexts are generally highly-trained on these tasks.
When these users engage with this type of advice, they frequently report that the system highlights cases that are irrelevant or that they are already aware of, causing them more work rather than less~\cite{Ginestra2019,kamran_evaluation_2024}. 
In other words, conventional AI tools may be of less use in straightforward cases where they are most accurate, leaving users in the dark on more complex decisions in which they \textit{want} support.

Since AI designs that automate aspects of an expert decision often result in low uptake~\cite{goh_large_2024,Ginestra2019,sivaraman_ignore_2023} or misalignments in goals and values~\cite{kawakami_improving_2022}, a growing number of AI systems have been designed to instead support decision-makers' \textit{reasoning} as they evaluate a case.
This literature aims to integrate AI information into decision-making in other ways beyond acceptance or rejection of an automated decision, such as ``negotiation'' ~\cite{sivaraman_ignore_2023}, consideration of clinical uncertainty~\cite{zhang_rethinking_2024}, or deliberation between multiple options~\cite{ma_towards_2025,reicherts_ai_2025}.
However, the actual process by which an AI system influences reasoning remains unclear, making it difficult to design new AI systems and predict how they might impact decisions.

Our work addresses this knowledge gap by developing a framework for human-AI decision-making centered around \textbf{intelligent reasoning cues}: discrete pieces of AI-derived information that can exert influence during the process of making a decision.
In our framework, AI interfaces can have multiple reasoning cues embedded within them, each comprising information that might support a decision-maker in forming an assessment or plan for a situation.
Unlike the ``second-opinion'' model, where the user must undertake the cognitive burden of evaluating the AI's advice and deciding whether to rely on it, users of intelligent reasoning cues find \textit{value} in a reasoning cue if it progresses their thinking towards a decision.
The field of HCI has already made progress in designing new types of intelligent reasoning cues using concepts such as ``Unremarkable AI''~\cite{Yang2019}, ``forward reasoning support''~\cite{zhang_forward_2021}, and others~\cite{kawakami_why_2022,zhang_rethinking_2024}. 
There is also potential to create even more adaptive cues with generative AI (GenAI)~\cite{ma_towards_2025,reicherts_ai_2025,yang_harnessing_2023}. 
Our framework seeks to explain the success of these emerging tools while pointing to design directions for even more effective reasoning cues.

To build our understanding of human-AI decision-making using our framework, we conducted a case study in the intensive care unit (ICU), a medical area ripe with potential for AI tools to support reasoning.
We focused on sepsis, a life-threatening condition in which the body's response to an infection injures its own tissues~\cite{CentersforDiseaseControlandPrevention2021}.
Sepsis can present in a wide variety of ways, requires time-sensitive and personalized decisions~\cite{Evans2021,Stoneking2011}. In addition, decision-making for sepsis remains suboptimal even among experts, leading to preventable morbidity and mortality~\cite{Stoneking2011}. 
Most ML-based tools for sepsis treatment can be thought of as offering second opinions -- they recommend optimal treatment plans given a patient's vital signs, laboratory values, and other clinical indicators~\cite{Komorowski2018,Peng2018,nanayakkara_unifying_2022}. 
However, despite their promising algorithmic performance, none of these systems have been adopted into practice due to concerns of bias~\cite{park_how_2024,Jeter2019} and incompatibility with experts' current decision strategies~\cite{sivaraman_ignore_2023}. 
As a result, these tools have failed to achieve their promise of improving patient outcomes.
Therefore, in this study we explored how a wider range of reasoning cues beyond recommendations might lead to different influences on ICU clinicians' decisions and perceptions of AI support.
The following research questions guided our work:
\begin{enumerate}[label={\bfseries RQ\arabic*.}, ref={\bfseries RQ\arabic*},itemsep=1ex]
    \item How do different intelligent reasoning cues influence the reasoning processes of ICU clinicians during treatment decisions? \label{rq:reasoning-influence}
    \item When do clinicians perceive value and usefulness in intelligent reasoning cues? \label{rq:perception}
\end{enumerate}

We first designed and developed eight intelligent reasoning cues -- such as unusual features about a case, insights into peer actions, or plan recommendations -- that have shown promise in prior studies of sepsis treatment and other decision-making tasks.
Different combinations of these reasoning cues were embedded in several AI-powered interfaces that could serve as decision support tools.
We then conducted two studies to understand how clinicians perceive and would be influenced by our intelligent reasoning cues: a series of contextual inquiries and semi-structured interviews with six ICU care teams, and a controlled think-aloud study in which 25 expert physicians made decisions on complex sepsis cases using the AI interfaces.

Our results demonstrate that reasoning cues are associated with specific \textit{patterns of influence} on the clinical decision-making process, such as Resolving Contradictions, Considering Alternatives, identifying a Plan Preference, and others.
We also find that clinicians perceive varying levels of value in individual cues depending on whether they support the \textit{right} aspects of a decision, how \textit{compatible} they are with current goals, and whether they are \textit{complementary} to their prior knowledge.
These findings can help us understand why some reasoning cues embedded in AI decision support systems are effective and others are not, especially when the cues were originally intended to calibrate reliance (such as AI explanations).
They also establish a path toward creating new intelligent reasoning cues that are more adaptive to individual case contexts, providing more value to decision-makers and positively influencing decisions.

\section{Related Work}

Our framework builds upon extensive HCI research on AI-based systems for decision-making (Sec. \ref{sec:human-ai-interaction}) as well as the literature on reasoning processes from psychology and decision science (Sec. \ref{sec:clinical-reasoning}).
This section situates our work among literature in human-AI decision-making generally, while Section \ref{sec:study-context} provides context specific to our case study on sepsis treatment.

\subsection{Human-AI Interaction for Decision Support}
\label{sec:human-ai-interaction}

Although supporting expert decisions has long been a central goal of automation and artificial intelligence research, the focus of decision support research has shifted as the underlying technologies have changed.
Early work, often using rule-based systems designed by experts~\cite{shortliffe_mycin_1977}, emphasized understanding and tailoring the level of automation needed for a particular task, ranging from simple data acquisition to fully autonomous decision-making~\cite{parasuraman_model_2000}. 
In clinical contexts, for instance, some decision support systems incorporated ergonomics-informed approaches such as Decision-Centered Design to align system design with the cognitive challenges faced by clinicians~\cite{schnittker_decision-centred_2019}.
Despite these systems' limited ability to deliver individualized insights, they did capture a broad range of possible roles for automation in a decision context~\cite{cummings_collaborative_2009}.

With the rise of modern machine learning, research has shifted toward understanding and calibrating humans' \textit{reliance} on AI outputs in making decisions~\cite{Lai2021,schemmer_appropriate_2023}. 
Driven by ML models' promising (but not perfect) performance on decision tasks, considerable literature has explored whether humans can perform these tasks better with ML advice, across domains such as child welfare \cite{kawakami_improving_2022}, aviation \cite{zhang_resilience_2023}, and health care \cite{Tschandl2020, jussupow_augmenting_2021, Gaube2021, jacobs_how_2021, panigutti_understanding_2022, prinster_care_2024, niraula_intricacies_2025}.  %
Often using simplified decision environments with finite choices, studies have evaluated whether model explanations~\cite{panigutti_understanding_2022,prinster_care_2024}, uncertainty estimates~\cite{Schaekermann2020}, user control~\cite{cheng_overcoming_2023}, or other interventions might promote reliance on correct advice and mitigate over-reliance on incorrect advice.
The AI information types shown in these studies, such as saliency maps, counterfactual explanations, and behavior descriptions, often show promising benefits for decision-making in individual scenarios~\cite{reverberi_experimental_2022,mothilal_explaining_2020,cabrera_improving_2023}, and the present work does not undermine their potential.
However, their effects on reliance and task performance are often mixed and inconsistent when compared across studies and contexts~\cite{vaccaro_when_2024}, suggesting that the quality of the human-AI team is dependent on a larger set of contextual factors.
By examining the use of AI information at the level of reasoning, we can better understand when these design elements are most useful to experts.

A few recent HCI works have begun to offer alternative ways of thinking about humans' use of AI-provided information, moving beyond appropriate reliance.
For example, \citeauthor{zhang_resilience_2023} use \textit{appropriation} as a guiding concept to explain how pilots might gain value from an intelligent system despite knowing that it cannot perfectly capture the nuances of their decision process~\cite{zhang_resilience_2023}.
Meanwhile, studies by \citeauthor{reicherts_ai_2025} and \citeauthor{ma_towards_2025} have leveraged the more flexible nature of GenAI to prototype decision support tools that help users \textit{evaluate} alternatives rather than simply recommending one~\cite{reicherts_ai_2025,ma_towards_2025}.
These studies suggest that rather than solely conceptualizing AI as a second opinion, a broader view of human-AI collaboration in decision-making is needed~\cite{gomez_human-ai_2025}. 
However, to our knowledge there is no model of human-AI interaction that can explain \textit{how} these new designs more productively influence users.
The primary goal of this work is to develop such a model, helping build more generalizable principles about what AI designs best support decision-making in different contexts.




\subsection{Theories of Reasoning in Decision-Making}
\label{sec:clinical-reasoning}

Understanding how experts make decisions with AI requires grounding in cognitive psychology, which has developed several frameworks of reasoning that apply to medicine as well as other decision-making domains. 
For example, \citeauthor{elstein_medical_1978}'s hypothetico-deductive model conceptualizes diagnostic clinical reasoning as a general cognitive skill in which clinicians generate a limited number of hypotheses, then selectively gather information to confirm or rule them out~\cite{elstein_medical_1978}. 
While this approach often serves as an idealized model in educational settings for how clinical reasoning should proceed~\cite{yazdani_five_2019}, other models may be more descriptive of reasoning in real-world and time-limited scenarios.
For instance, the illness script theory argues that experienced clinicians use internal structured representations of existing knowledge to identify representations of disease cases~\cite{custers_role_1998}, while the recognition-primed decision model explains that experts arrive at a solution by recognizing patterns in a decision context and evaluating possible plans one at a time~\cite{klein_recognition_1993}. 
Dual Process Theory serves as a useful alternative lens to understand clinical decision-making by distinguishing fast, intuitive, and heuristic-based reasoning (System 1) from slower, analytical, and effortful processes (System 2) \cite{norman_research_2005}.
These adaptations to high-stakes, time-pressured environments may explain why traditional recommendation-based models can fail to influence experts' decisions, especially if they only provide a second opinion after the human has formed a judgment~\cite{jussupow_augmenting_2021}.

As AI-based decision support has become more widespread, some recent studies have argued that these systems should focus on supporting reasoning rather than just the final decision~\cite{zhang_rethinking_2024,sokol_artificial_2025,VanBaalen2021}.
\citeauthor{medlock_modeling_2016} proposed a two-stream framework describing how a clinical decision support system must optimize both the clinical advice and the presentation of that advice, based on the patient and the clinician's cognitive context~\cite{medlock_modeling_2016}.
However, frameworks such as this are not well integrated with models of human reasoning~\cite{greenes_clinical_2018}, making it challenging to pinpoint the role of different aspects of the system's design on users' reasoning processes. 
Importantly, modern clinical decision support frameworks offer limited understanding of where specific pieces of external information enter and shape clinicians’ reasoning processes.
They also assume that the decision support system's outputs are static functions of the case data, as opposed to the emerging interactive designs built on modern machine learning models. 

Meanwhile, distributed cognition theory and other holistic analysis methods in HCI emphasize that interactive systems form an integral part of people's cognitive processes during complex tasks~\cite{hollan_distributed_2000}. 
Yet existing HCI theory does not yet explain how AI information, often less predictable and more discordant with human intuition, integrates into the reasoning process.
This further emphasizes the need to study not only how expert decision-makers reason, but also how different forms of AI-derived information interact with and alter that reasoning.

\section{Proposed Framework: Intelligent Reasoning Cues}

Our framework is built on a small but significant expansion of how decisions with AI are typically conceptualized.
In the conventional model that underpins the majority of human-AI decision-making literature~\cite{lee_trust_2004,arnold_differential_2006,Lai2021}, the AI interface is thought of as presenting a recommendation or prediction that the user must accept or reject (Fig. \ref{fig:cognitive-models}a).
If other information is provided, such as an explanation, it is usually designed or evaluated with the intent to calibrate a user's reliance on the initial piece of information~\cite{papenmeier_its_2022,schemmer_appropriate_2023}.
However, HCI literature has shown that decision-makers' successful use of AI may hinge on factors \textit{beyond} appropriate reliance~\cite{sivaraman_ignore_2023,kawakami_improving_2022,zhang_resilience_2023}, a likely reason for the mixed or null results across many explainable AI studies~\cite{vaccaro_when_2024}.
Moreover, emerging AI-based decision support systems have increasingly and successfully aimed to support users in ways beyond acceptance or rejection, such as prompting information-gathering~\cite{zhang_rethinking_2024}, weighing multiple alternatives~\cite{reicherts_ai_2025,ma_towards_2025}, or exploring counterfactuals~\cite{mothilal_explaining_2020}.
How can we better understand the diverse ways that these tools can integrate into decision-making?

We propose \textbf{intelligent reasoning cues} as a new framework allowing for a broader range of AI roles than existing frameworks, making it a useful lens to understand human-AI decision-making.
In our model (shown in Fig. \ref{fig:cognitive-models}b), AI-based decision support interfaces (hereafter referred to as ``AI interfaces'') are seen as collections of ``reasoning cues'' that can influence a decision-maker's thought process.
Unlike the conventional model, where the principal role of AI advice is solely to be accepted by the user in favor of their own judgment~\cite{schemmer_appropriate_2023}, this model predicts that reasoning cues will have open-ended but consistent \textit{patterns of influence} on the human's thought process.
Moreover, the human's interaction with the AI interface can trigger the underlying model to produce new reasoning cues, as can be seen in more recent adaptive designs ~\cite{reicherts_ai_2025,zhang_rethinking_2024}.
Our conceptualization therefore builds upon earlier models for human interaction with automation~\cite{parasuraman_model_2000} and classical decision support systems~\cite{hollnagel_information_1987}, while accounting for the interactivity and adaptiveness of modern AI interfaces.


\begin{figure}
    \centering
    \includegraphics[width=\linewidth]{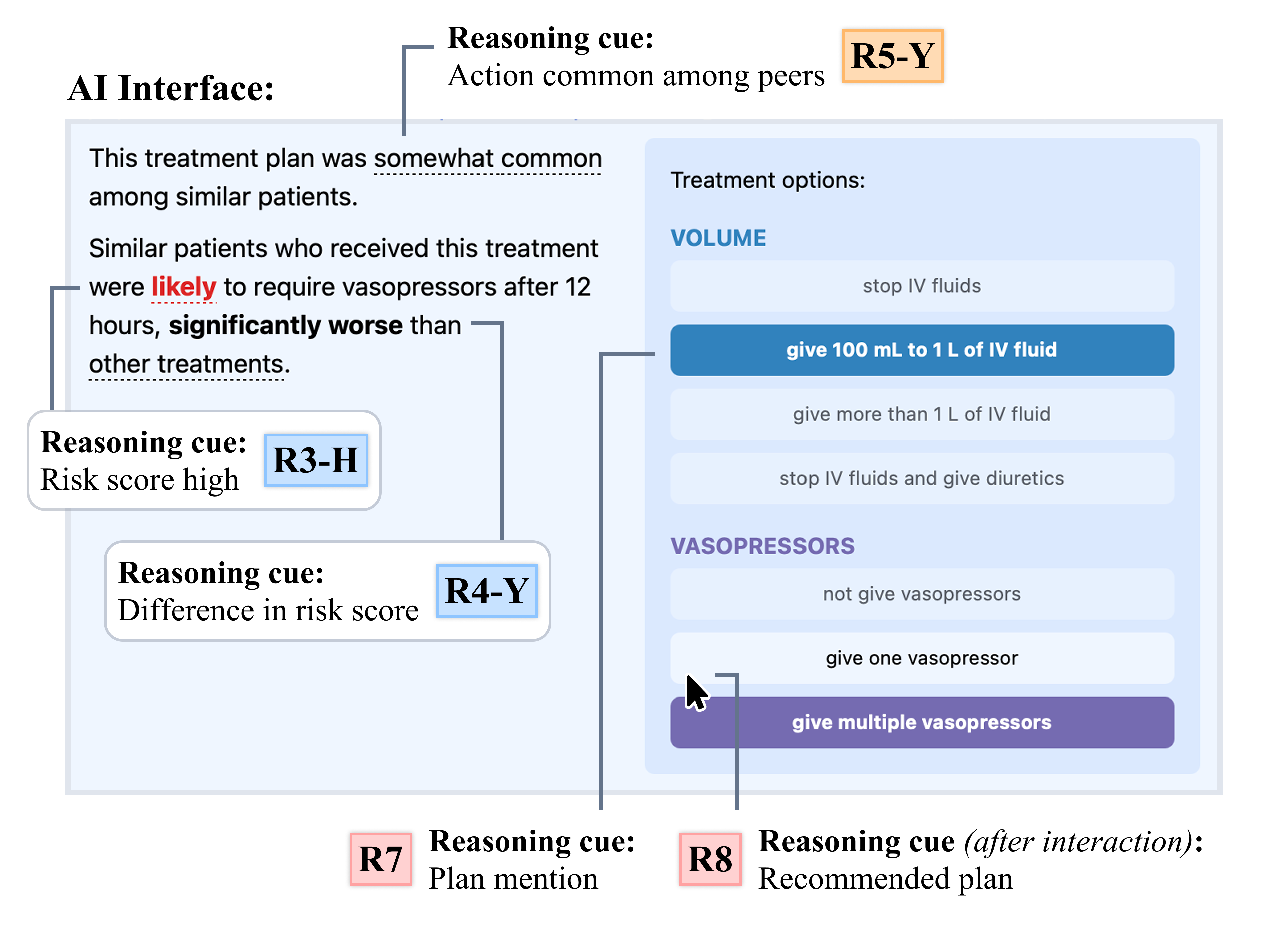}
    \caption{Our framework models AI interfaces as collections of multiple intelligent reasoning cues, shown here on the Interactive Treatment Risk interface from our study. Each reasoning cue (labeled by codes R1-R8 and described in Sec. \ref{sec:reasoning-cue-desc}) conveys to the user a discrete insight that could support a decision.}
    \label{fig:reasoning-cue-preview}
\end{figure}

To demonstrate the relationship between intelligent reasoning cues and AI interfaces, let us consider the Interactive Treatment Risk prototype from our study (shown in Fig. \ref{fig:reasoning-cue-preview} and described further in Sec. \ref{sec:reasoning-cue-desc}), which depicts the risk of a patient with sepsis requiring second-line drug conditional on a plan selected by the user.
This system contains several elements that constitute reasoning cues: a prediction about the patient's risk of needing the drug, the frequency of the selected plan being taken among similar patients, and others. 
Note that the right side of the interface (which shows a series of plan mentions) is static, so it can influence the user but cannot be influenced \textit{by} the user; however, the other cues can change depending on which plan the user selects.
Interactive interfaces can be more powerful by adapting the AI information to a user's needs, thus influencing decision-making without necessarily making prescriptive recommendations. 
However, they can also make it more difficult to pinpoint the specific influence of the AI on reasoning and determine its impact on decisional outcomes.
By conceptualizing the interface in terms of changing reasoning cues, we can better explain why particular interactions might lead to shifts in behavior.

\section{Study Context: Sepsis Treatment}
\label{sec:study-context}

We used the intelligent reasoning cues framework described above to develop a case study investigating how AI information affects reasoning about sepsis treatment decisions.

Sepsis is a condition in which the body responds to infection in a dysregulated way, resulting in organ dysfunction~\cite{Evans2021}.
As sepsis is a leading cause of mortality in hospitals~\cite{CentersforDiseaseControlandPrevention2021}, improving its management is of utmost importance for clinicians and policy makers. 
Clinical guidelines for sepsis are based on both prospective randomized clinical trials and real-world observational studies showing that, among the universe of plausible treatment strategies, some treatment strategies are associated with lower patient mortality than others~\cite{Evans2021}. 
In this way, some decisions are empirically better than other decisions, though this can't necessarily be known by a human decision-maker at the time of the decision. 
An essential group of treatment strategies centers around hemodynamic resuscitation, the process of correcting low blood pressure (hypotension) and restoring organ perfusion using intravenous (IV) fluids and/or vasopressors (e.g., norepinephrine). 
Later in a sepsis patient's clinical course, it may also be necessary to remove fluid previously given through a process known as diuresis, preventing potential organ injury by fluid overload. 
These three treatments -- \textbf{IV fluids}, \textbf{vasopressors}, and \textbf{diuretics} -- are the main treatments considered in our study.

Although some treatment guidelines exist for sepsis, considerable clinical uncertainty remains around how fluids, vasopressors, and diuretics should be administered to maximize survival. 
All three treatments carry potential benefits as well as risks that can prolong hospital stay and necessitate more intensive care. 
For example, excess fluid can result in ``volume overload'' or accumulation of fluid in the lungs~\cite{Marik2020,Evans2021},
and it can be difficult to determine whether to liberally administer fluids followed by diuresis or to be more restrictive initially~\cite{the_national_heart_lung_and_blood_institute_prevention_and_early_treatment_of_acute_lung_injury_clinical_trials_network_early_2023}.
Similarly, vasopressors can contribute to irregular heart rhythms that can cause further complications, and it is unclear whether to administer these drugs earlier or to favor fluids first~\cite{Evans2021}. 
Treatment decisions can be even further complicated by the presence of comorbidities such as heart failure or kidney injury~\cite{munroe_understanding_2024}, which not only increase mortality risk but can limit the use of fluids and vasporessors that are essential sepsis treatments~\cite{jones_sepsis_2021}.
The inability of clinical guidelines to perfectly prescribe the best decision for every patient is the central challenge of treatment decision-making in sepsis.


In response to these challenges, a growing body of research has focused on developing AI systems to support more proactive and patient-specific decision-making. 
Much of this work has emphasized early \textit{identification and diagnosis} of sepsis in a patient's course, as this can significantly reduce their mortality risk~\cite{Sendak2020,adams_prospective_2022,zhang_rethinking_2024}. Although potentially promising, these efforts have led to only modest improvements in patient outcomes, largely because they don't consistently lead to behavior change among clinicians~\cite{Ginestra2019,kamran_evaluation_2024}.
Other efforts, meanwhile, have aimed to recommend optimal treatment strategies \textit{after} a sepsis diagnosis, often using reinforcement learning on large observational datasets~\cite{Komorowski2018,Peng2018,nanayakkara_unifying_2022}.
Early treatment recommendation models were critiqued for recommending nonsensical actions and systematically under-treating patients with more severe sepsis~\cite{Jeter2019}, a finding underscored by qualitative work showing these models to ICU clinicians~\cite{sivaraman_ignore_2023}.
Although the underlying modeling techniques have since improved, it remains unclear whether treatment recommendations based on present-day observational datasets can ever lead to better outcomes than clinicians~\cite{park_how_2024}.
Rather than giving a recommendation that users must either agree or disagree with, designing more indirect tools for clinicians' intermediate reasoning stages may be a promising way to improve decisions when recommendations are hard to get right.


\section{Study Methods}

To explore how clinicians might be influenced by and perceive different intelligent reasoning cues for sepsis treatment decisions, we developed eight reasoning cue designs and a series of AI interfaces based on prior systems developed for sepsis and other expert contexts~\cite{sivaraman_ignore_2023,kalimouttou_optimal_2025,yong_deep_2024,huang_reinforcement_2022}.
Then we systematically evaluated clinicians' reasoning on real sepsis cases through two complementary studies: a series of contextual inquiries and interviews with critical care teams (\textbf{Study 1}) and a think-aloud study with critical care physicians (\textbf{Study 2}).
Below, we describe methods for designing the AI interfaces with intelligent reasoning cues, conducting the two studies, and analyzing the results.

\subsection{AI System Design}


We aimed to create AI interfaces that could plausibly be deployed in real life, using information that would be available in a modern ICU electronic health record. 
We considered developing multiple ML models tailored to the predictive tasks needed for each reasoning cue, but were concerned that this could result in inconsistent behavior due to chance differences in optimization.
Therefore, we trained a single \textit{deep transformer-based encoder} to learn similarities in patient states, then used the nearest neighbors that it identified for a given case to populate our intelligent reasoning cues.

\subsubsection{Data and Model}
The machine learning model was trained and evaluated using the widely-used MIMIC-IV dataset~\cite{johnson2020mimic}, which contains high-density electronic health record data on ICU patients at a large academic medical center in the northeastern United States.
Patients who had suspected sepsis (measured by the presence of antibiotic administration and microbial cultures taken within 24 hours of each other) and who were in the ICU for at least 12 hours were included.
Any data past the 95th percentile of ICU stay length (about 17 days) was removed.
Consistent with prior work~\cite{Komorowski2018}, we also removed timesteps that occurred after comfort measures only (CMO) orders were enacted for a patient, as this would typically signify that the treating clinicians' intent was no longer to optimize the patient's chance of survival.
This resulted in a total of 30,398 patients with an average of 5 days of data per patient; half of these patients were used for training while the other half were used for evaluation and case selection for the study.

Each patient was modeled as a series of timesteps containing their demographics, prior diagnosis history, vital signs, most recent labs, and treatments at four-hour intervals.
To train the model, we used a denoising autoencoder task in which the model took corrupted versions of the patient trajectory as input and attempted to predict the original version at each timestep.
We chose an autoencoder model since this would allow us to represent patients in a high-dimensional embedding space, where similar patient states (and ideally patients with similar treatments and outcomes) would be clustered closer together.
Using this patient representation, we could generate a wide range of reasoning cues for a new patient using information about the \textit{nearest neighbors} of that patient in the embedding space.
The data processing and modeling code is open-source\footnote{\url{https://github.com/cmudig/sepsis-reasoning-dashboard}}.


\subsubsection{Intelligent Reasoning Cues}
\label{sec:reasoning-cue-desc}
The space of possible intelligent reasoning cues is naturally large, and enumerating all types of cues is beyond the scope of this paper.
Instead, our study included eight intelligent reasoning cues, that covered the major information types used in prior AI-based decision support tools for sepsis treatment~\cite{sivaraman_ignore_2023,hyland_early_2020,kalimouttou_optimal_2025,Komorowski2018}.
We also included cues based on explainable AI designs from other analogous contexts, such as unusual features or conditional differences in risk previously used to support child welfare decisions~\cite{kawakami_why_2022,zytek_sibyl_2022}.
Rather than exhaustively testing a larger set of cues, this selection allowed us to compare approaches that were already known to have potential utility for our decision task.
The reasoning cues we created are listed in Fig. \ref{fig:reasoning-cue-interfaces}a and fell into four main categories:

\paragraph{Case description cues.}
\begin{itemize}
    \item \consistentfeatures{} \consistentfeatures*{}  showed up to three features in the given patient that were most consistent among their 100 nearest neighbors, similar to a feature explanation. 
    \item \unusualfeatures{} \unusualfeatures*{} showed the features that were most different from a case's neighbors, identified the same way as above.
\end{itemize}
\paragraph{Risk prediction cues.}
\begin{itemize}
    \item \anyrisk{} \anyrisk*{} displayed the probability of a given event among the case's nearest neighbors. Because different values of the risk score could induce different forms of reasoning, we denote extreme (low/high) risk scores with the cue \lowhighrisk{} and moderate risk scores with \moderaterisk{}. 
    \item \diffrisk{} \diffrisk*{} indicated whether similar patients who received a particular treatment plan had a different risk than other similar patients (\diffriskyes{} if there was a difference, \diffriskno{} otherwise).
\end{itemize}
\paragraph{Peer action cues.}
\begin{itemize}
    \item \commonaction{} \commonaction*{} (\commonactionyes{} and \commonactionno{}) indicated how frequently a treatment plan was taken for similar patients.
    \item \consensusaction{} \consensusaction*{} (\consensusactionyes{} and \consensusactionno{}) showed whether there was or was not a single plan that was consistently taken (observed in more than 60\% of similar cases).
\end{itemize}
\paragraph{Plan cues.}
\begin{itemize}
    \item \planmention{} \planmention*{} was a cue simply referring to the visual existence of a treatment plan on the interface. 
    \item \recommendedplan{} \recommendedplan*{} represented a plan that was positioned as ``best'' in some way, either by peer actions or by risk scores.
\end{itemize}

\subsubsection{AI Interfaces}

Each AI interface used in our study incorporated one or more of the reasoning cues above.
It was necessary to combine reasoning cues to create our interfaces because many reasoning cues build on each other; for instance, it would be impossible to show a cue such as \diffrisk{} (plan-dependent risk) without including a plan mention (\planmention{}).
Moreover, testing all possible combinations of even this small set of reasoning cues would be infeasible.

We started by creating designs based on existing decision support tools for sepsis, such as mortality risk scores~\cite{gao_prediction_2024} and treatment recommendations~\cite{Komorowski2018}.
Then we drew from recent HCI literature~\cite{kawakami_why_2022,zytek_sibyl_2022,zhang_resilience_2023,zhang_rethinking_2024} to create additional interfaces that provided more interactivity and combined more reasoning cues together.
The resulting seven designs were implemented in a web interface using the data and model described above, after which we refined their clarity and usefulness by soliciting feedback from a sepsis expert.
As shown in Fig. \ref{fig:reasoning-cue-interfaces}b, the interfaces we created were as follows:


\paragraph{Case Features.} Similar to a feature explanation~\cite{Lundberg2017}, this interface was designed as a descriptive tool to understand the salient features of the current case, incorporating those that were most consistent (\consistentfeatures{}) and most unusual (\unusualfeatures{}) compared to similar cases. 
\paragraph{Treatment Risk/Mortality Risk.} These two interfaces depicted the overall likelihood of the patient having a certain outcome (\anyrisk{}) regardless of the treatment action, similar to existing risk assessment tools~\cite{gao_prediction_2024}. To explore whether different predictive targets influenced the amount of value clinicians got out of risk scores, the two variants incorporated different outcomes: the risk of the patient being given vasopressors after 12 hours, a proxy for their future disease severity; and the risk of mortality within the current admission.
\paragraph{Interactive Treatment Risk/Mortality Risk.} This pair of interfaces was analogous to the preceding ones but incorporated more reasoning cues through interactivity. They allowed the user to select a treatment plan in terms of volume (IV fluids and diuretics) and vasopressors, constituting a plan mention (\planmention{}). Once a treatment plan was selected, the system showed a risk score (\anyrisk{}) conditional on the patient receiving that treatment. The interfaces also highlighted if there was a statistically significant difference in the risk score for the selected treatment compared to the others (\diffrisk{}), making it possible for the clinician to infer a recommended plan (\recommendedplan{}) when the risk score was significantly better than other treatments. These interfaces also depicted how frequently the selected treatment was taken for similar patients (\commonaction{}), allowing the clinician to estimate the confidence in the prediction while giving a sense of the prior clinicians' actions. When a plan was taken in fewer than 10 out of 100 similar patients, no recommendation was shown (\commonactionno{}).
\paragraph{Prior Clinician Actions.} This interface summarized the action frequencies within the same categories (volume and vasopressors) without showing any outcome data. Therefore, this interface always presented plan mentions (\planmention{}) and cues related to the frequency of an action (\commonaction{}). It also noted whether there was a single action that was most commonly taken (\consensusactionyes{}), or whether clinicians were mostly inconsistent for similar cases (\consensusactionno{}). Although not explicit, the actions most commonly taken in each category could be considered a recommendation (\recommendedplan{}).
\paragraph{Treatment Recommendation.} Finally, we created an interface that made a direct treatment recommendation based on the model output (both a plan mention \planmention{} and a recommendation \recommendedplan{}). This interface was the simplest, most direct, and most consistent with conventional AI-based decision support systems. 

\subsection{Study 1: Contextual Inquiry with Critical Care Teams}

The first study allowed us to observe clinical decision-making in a real-world setting, often difficult to capture in controlled experiments on clinical reasoning~\cite{mansoori_cognitive_2025,munroe_understanding_2024,sivaraman_ignore_2023}.
In this stage, we aimed to deeply understand clinicians' existing reasoning processes around sepsis \textit{before} the introduction of any intelligent reasoning cues, so that experts could share more open-ended perspectives on AI-based information.

Two members of our research team joined six critical care teams during their morning rounds in ICUs in four different hospitals within a large academic hospital system. 
Typical of most academic ICU teams, these groups usually consisted of 4--10 individuals, including an attending physician, one or more clinical fellows and residents, a clinical pharmacist, as well as respiratory therapists and bedside nurses. 
Some teams also included an advanced practice provider (APP), who handled many care decisions alongside the attendings and fellows.
During the rounds, the team visited each patient that was in the ICU that day (typically about 5--10 patients reviewed within a 2-hour period).
The bedside nurses and trainees gave detailed reports on each patient's ICU course and current status, then provided recommendations that were discussed with the group and approved by the attending or APP.
Our focus was on understanding the challenges and uncertainties that these decision-makers faced around managing patient care, particularly for patients with suspected or confirmed sepsis.

Following a standard contextual inquiry approach~\cite{holtzblatt_principles_2017}, we observed each morning rounds and took detailed notes on the ongoing conversations and patient care decisions, asking questions related to our focus area when possible.
After each set of rounds, we invited one or more of the clinicians we observed for a semi-structured interview, which usually lasted between 10 and 20 minutes depending on clinicians' ICU schedules. 
During this interview we followed up on the most challenging cases of the day and probed how they might envision AI supporting their reasoning.
When time permitted, we also asked interviewees to walk us through how they would make a treatment decision on two patient vignettes, giving us a reference point to understand the thought processes in Study 2 and to ground discussions about AI support.
In addition to the six interviews following each contextual inquiry, we also conducted one separate interview with an expert attending clinician.
With participants' consent and IRB approval, these interviews were audio-recorded and transcribed for analysis.

\subsection{Study 2: Think-Aloud Sessions}

In the second study, we collected detailed examples of reasoning on a consistent set of patients while varying the types of available AI information using intelligent reasoning cues.
Our use of working AI interfaces in this study allowed clinicians to provide more grounded feedback on our specific designs, complementing the more general feedback from Study 1.
We also collected this data with a population of clinical fellows, who have completed their medical degree and residency and are just beginning to make autonomous patient care decisions.
ICU fellows typically have experience with sepsis, but may still have uncertainty about complex cases and therefore be more likely to benefit from AI support~\cite{jacobs_how_2021}.


We recruited 25 fellows from the same health system as Study 1 as well as from other institutions around the United States.
Each clinician took part in a think-aloud decision-making study that took place on Zoom and lasted 30--45 minutes, using a web-based tool we developed for the study (shown in Fig. \ref{fig:study-interface} in Appendix \ref{app:study-2-details}). Similar to prior think-aloud studies in health care~\cite{sivaraman_ignore_2023,anjara_examining_2023}, participants were first familiarized with the study interface, asked to make decisions on four patient cases aided by a variety of reasoning cues, and finally asked a set of semi-structured debriefing questions.
They were compensated 75 USD for their participation, and the sessions were video and screen-recorded with the participants' consent and IRB approval.

\begin{table*}[]
    \centering
    \begin{tabular}{p{2.3cm}|p{2.8cm}p{2.8cm}p{2.8cm}p{2.8cm}}
    \toprule
\textbf{Pseudonym} & \textbf{Deborah Walker} & \textbf{Heather Thompson} & \textbf{Richard Sullivan} & \textbf{Esther Goldberg} \\
\midrule
\textbf{Profile} & Female, age 60s; hour 8 of ICU stay & Female, age 50s; hour 8 of ICU stay & Male, age 70s; hour 52 of ICU stay & Female, age 80s; hour 20 of ICU stay \\
\textbf{Complicating Factors} & Low blood pressure while already on vasopressors; conflicting indicators of responsiveness to IV fluid; low sodium & Low blood pressure could be caused by sepsis and/or liver failure; no urine output since admission & Evaluating later in ICU course; increasing oxygen requirements possibly due to fluid overload; receiving multiple vasopressors & Elderly but at risk of needing invasive ventilation if given IV fluids; new-onset heart arrhythmia \\
\textbf{Original \newline Clinician \newline Decision} & Give > 1 L of fluids, add a second vasopressor & Give 100 mL - 1 L of fluids, no vasopressor & Give no fluids, continue multiple vasopressors & Give diuretics, no vasopressor\\ \bottomrule
\end{tabular}
    \caption{Summary of the patient cases used in Study 2. These patients were selected from the MIMIC-IV database and assigned pseudonyms for purposes of the study.}
    \label{tab:patient-profiles}
\end{table*}

\begin{figure*}
    \centering
    \includegraphics[width=\linewidth]{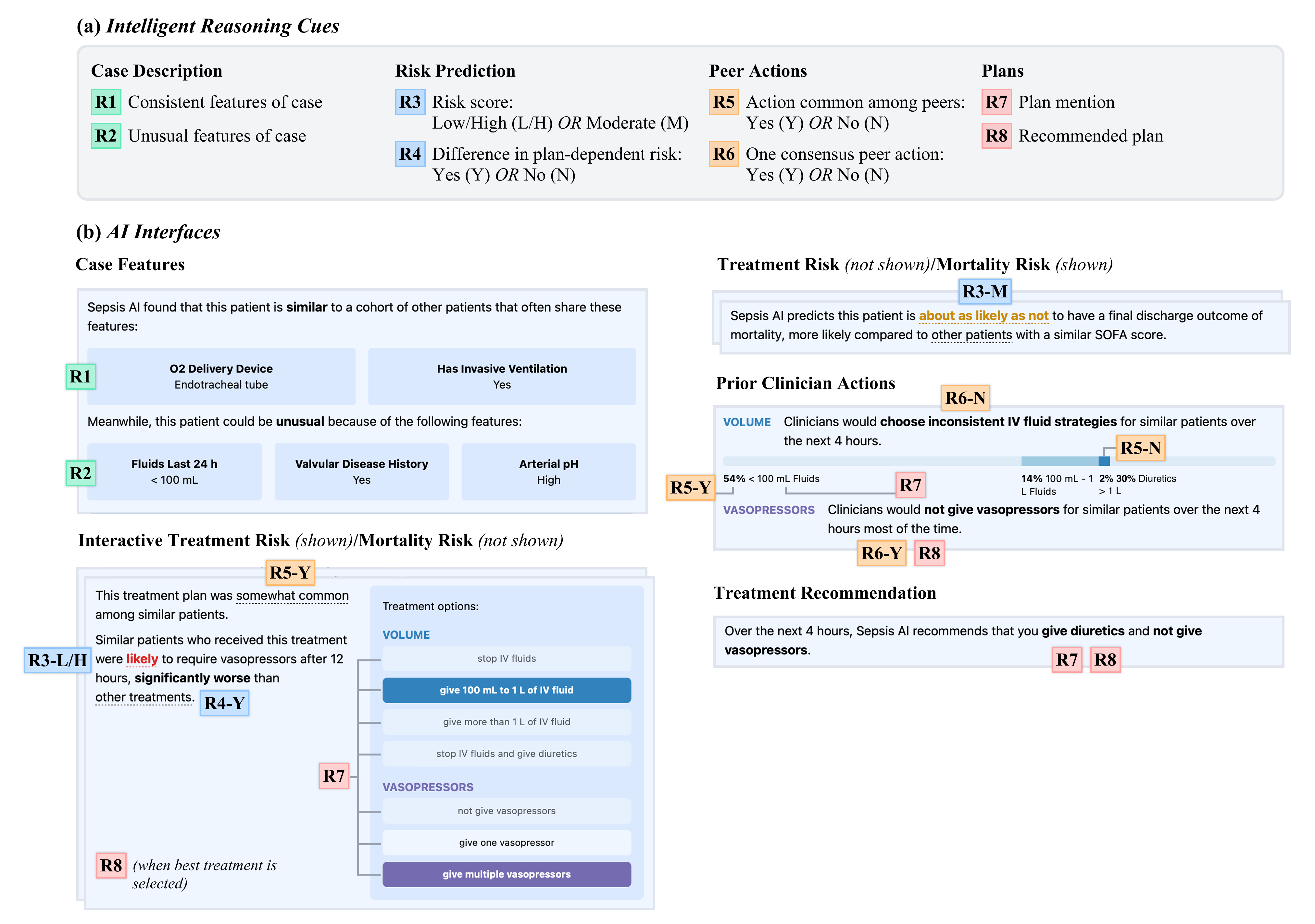}
    \caption{(a) \textbf{Intelligent reasoning cues}, numbered R1-R8, considered in this study. (b) \textbf{AI interfaces}, each containing multiple reasoning cues. Treatment Risk and Mortality Risk refer to different variants of the same interfaces with different predictive targets.}
    \label{fig:reasoning-cue-interfaces}
    \includegraphics[width=\linewidth]{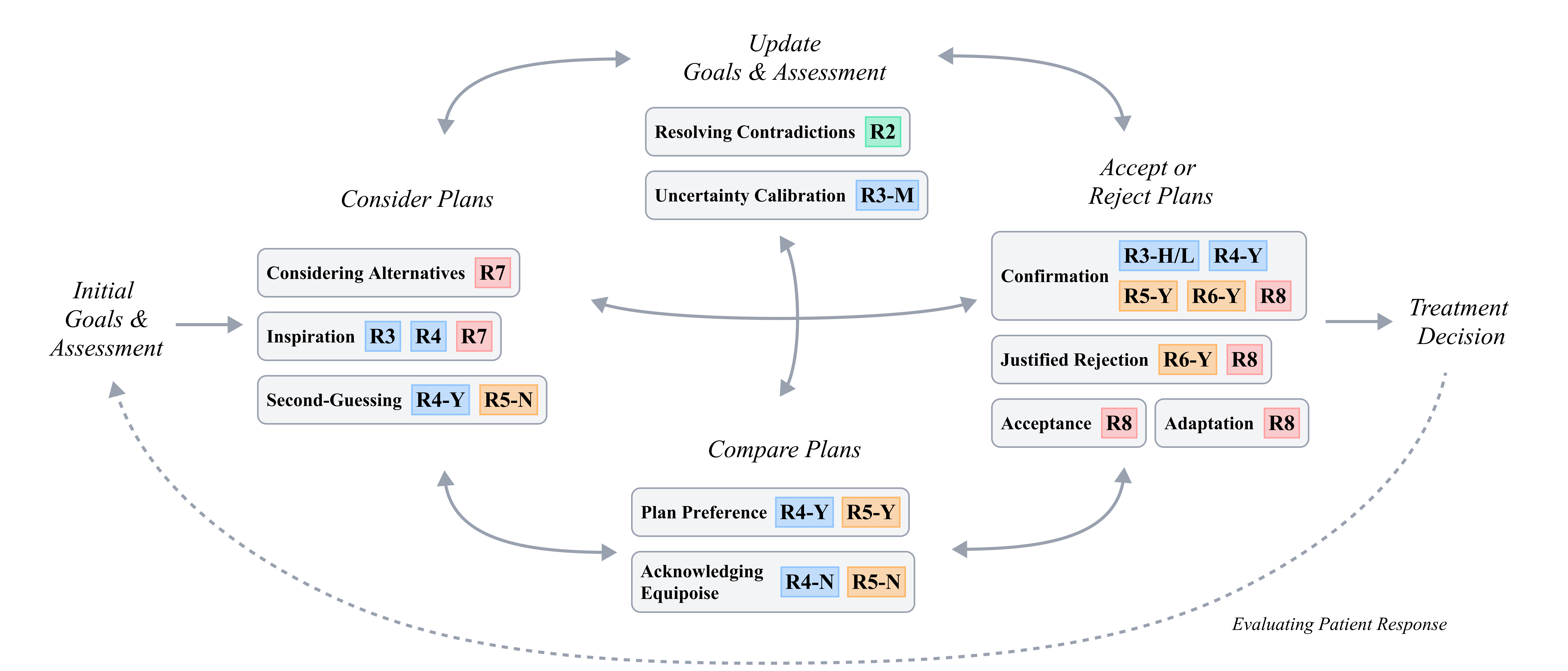}
    \caption{\textbf{Patterns of influence} (shaded boxes) of the intelligent reasoning cues as observed in our study, mapped onto a typical clinical-decision making workflow. Each pattern was most often observed after participants engaged with} specific reasoning cues denoted by their codes R1-R8.
    \label{fig:decision-workflow}
\end{figure*}

\paragraph{Relationship with Study 1}
While the formal analyses of both studies were conducted simultaneously and are thus presented together, the design of Study 2 was shaped by high-level insights we gained from team discussions following Study 1. 
For example, it was evident from Study 1 that clinicians most desired support in reasoning about complex cases, which often have competing treatment priorities beyond just sepsis (outlined in Sec. \ref{sec:study-context}). 
This motivated us to focus on complex patients as stimuli for the study, as described below. 
In addition, Study 1 showed that even across ICUs in the same hospital system, clinicians vary dramatically in their practice patterns, and they require different forms of reasoning assistance accordingly. 
As a result, we decided to create reasoning cues that encompassed a range of predictive targets and potential roles in reasoning. 
By conducting these two studies sequentially and then analyzing them together, we were able to refine the focus of Study 2 while incorporating complementary insights from real-world practice into our analysis.

\paragraph{Case Selection.}
Since all participants would be providing decisions on the same four cases, it was important to choose patients that were sufficiently complex and challenging to yield variation in decisions.
To narrow down from the thousands of available records in our MIMIC-IV evaluation set, we first filtered for patients who had abnormal measurements in several organ systems and who were considered unusual based on the transformer model's embedding.
Then, we searched for scenarios of ``classic'' complex sepsis patients provided by the critical care expert on our research team, such as ``patients with borderline blood pressure and a heart failure history who have recently received a lot of IV fluid.'' 
Finally, we selected four cases, listed in Table \ref{tab:patient-profiles}, that matched these scenarios and had interesting decision aspects based on their discharge notes.
We assigned each patient a pseudonym and wrote short vignettes to accompany the structured EHR data in our interface, which included demographics, illness history, vitals, labs, and prior treatments for the patient's ICU stay so far.
More information on what data was available for each patient is provided in Appendix \ref{app:study-interface}.

\paragraph{Randomization.}
Out of the four decisions made by each participant, three were made using an AI interface and one was made without AI.
The study used a within-subjects design in which participants had access to a different AI interface in each case, allowing them to make comparisons during the debrief session (although each person only saw three of the seven total interfaces).
Because fully crossing the interface and case variables would result in very few counts per interface per case, we opted to limit each AI interface to the two patient cases in which our model produced more complete reasoning cues (e.g., showing at least two unusual features for \unusualfeatures{} or at least two viable plans for a difference in risk \diffrisk{}).
This partially crossed design resulted in 5-6 observed decisions per case per AI interface, allowing us to identify consistent decision-making patterns even within a small sample size. 

\paragraph{Participant Workflow.}
Although it would be impossible to perfectly replicate participants' real workplace context, we aimed to integrate as much realism into the study procedure as possible (the full protocol is shown in Appendix \ref{app:study-protocol}).
Participants were told to imagine that they were preparing to discuss a series of patients at morning rounds, a common routine in the academic ICUs we observed in Study 1.
We asked participants to review each patient case and come up with a treatment recommendation to present during rounds, and to describe their thought processes to us as if we were trainees shadowing them.
They were also informed that they would see a series of different insights from an AI system for sepsis, and they were given a short summary of how the system works.
They could see the AI's outputs at any time while reviewing each case, and they were instructed to use the AI in any way they saw fit.
After reviewing the patient case and the AI output if one was provided, participants were asked to describe their decision verbally in terms of volume (IV fluids and diuretics) and vasopressors.
This open-ended approach allowed us to capture nuance in clinicians' thought process and decisions without restricting whether or when they might use the AI.

\subsection{Analysis}
\label{sec:analysis-method}

To answer \ref{rq:reasoning-influence}, the transcripts from participants' think-aloud sessions were coded by three researchers. 
We supplemented the transcripts with annotations about when participants engaged with reasoning cues and other parts of the interface.
To capture their reasoning process in detail, we coded when clinicians made a \textit{hypothesis} about the patient's status, expressed an \textit{uncertainty} about the patient's state, or stated an actionable \textit{plan} to treat the patient in terms of volume (IV fluids or diuretics) or vasopressors. 
We also tracked the \textit{final outcome} of a thought to note whether a hypothesis or plan was ultimately accepted, and described its role in a clinician's final decision. To assess AI influence on clinician decision-making, we tracked which \textit{reasoning cue} participants saw, and described the \textit{role of the AI} if it clearly influenced their thought. 
Each transcript was coded by at least two researchers, and conflicts were resolved through extensive discussion among the research team, which included experts in HCI, behavioral science, and medicine. This process allowed us to consolidate  each participant's complex decision-making into codes that capture individual thoughts while still preserving continuity in how thoughts contributed to final plans (see example in Table \ref{tab:interview-coding} in Appendix \ref{app:analysis-coding}).

We then conducted two types of affinity mapping on the resulting 544 codes. First, we grouped codes \textit{within} each patient case by looking for patterns associated with the lines of thinking for each patient. 
Second, we analyzed codes associated with AI use \textit{across} all patients to identify broader patterns in the interaction between reasoning cues and participants' decisions. 
This analysis enabled us to identify both the patterns of influence of reasoning cues on clinician thought processes and how such patterns manifested in specific patient cases.
(More information on our affinity mapping process is provided in Appendix \ref{app:affinity-mapping}.)


For \ref{rq:perception}, we used standard thematic analysis to analyze the notes and interview transcripts from Study 1 and responses to the debriefing questions in Study 2.
Open coding was performed by two researchers and disagreements were resolved by group discussion, resulting in 254 codes for Study 1 and 275 codes from Study 2.
These were then coalesced into 15 high-level themes, which we present in three categories in Sec. \ref{sec:results-perceptions}.

\section{Study Results}

The clinical decision-making processes we observed were complex, non-linear, and shaped by evolving information, changing priorities, and contextual constraints. 
While some clinicians were resistant to the idea of incorporating AI-based information into this process, we found that many participants \textit{did} see value in the intelligent reasoning cues embedded in our AI interfaces.
Our think-aloud study participants used these cues in diverse ways to help them interpret patient measurements and observations, assess the patient's overall state, and form and evaluate treatment plans.
Moreover, participants in both studies expressed detailed preferences on how useful or not different types of cues could be to them, underscoring the value of evaluating decision support systems in terms of their reasoning cues.

We first describe patterns in how the AI interacted with clinicians' reasoning (\ref{rq:reasoning-influence}) in Sec. \ref{sec:results-reasoning-patterns}, followed by the high-level themes in their perceptions of AI influence (\ref{rq:perception}) in Sec. \ref{sec:results-perceptions}.
Throughout these sections, we use codes such as C1, C2, etc. to refer to findings from the care teams in our contextual inquiry (Study 1), and P1, P2, etc. to denote participants in the think-aloud (Study 2).

\subsection{Patterns of Influence of Reasoning Cues}
\label{sec:results-reasoning-patterns}

To structure our results for \ref{rq:reasoning-influence}, we mapped the interactions with reasoning cues that we observed to the typical clinical reasoning workflow we observed across both studies. 
As shown in Fig. \ref{fig:decision-workflow}, this process typically began with establishing an initial assessment and goals of care, followed by iteration between multiple stages of thinking: \textit{Consider Plans}, \textit{Update Goals \& Assessment}, \textit{Accept or Reject Plans}, and \textit{Compare Plans}. 
This loop ultimately led to a treatment decision, the endpoint of each case in Study 2. 
Participants emphasized that in a real setting this decision-making process would not end at a treatment plan, but rather would also include when and how to evaluate the patient's response to inform subsequent assessments and goals. 

This workflow is grounded in the clinical reasoning theories discussed in Sec. \ref{sec:clinical-reasoning}, such as pattern recognition for forming an initial impression and the hypothetico-deductive approach for testing and revising that impression through cycles of assessment and plan evaluation. Our analysis extends these models by pinpointing where AI-derived cues shape these cycles, such as redirecting initial impressions or prompting reevaluation of emerging hypotheses. 



Our qualitative analysis of hypotheses, uncertainties, and plans identified eleven distinct \textit{patterns of influence} that intelligent reasoning cues enacted on participants' thinking, as described below.

\subsubsection{Considering Plans}
Any interface that mentioned a possible plan (\planmention{}) could lead a clinician to consider it, regardless of what information the AI provided about that plan.
A particularly consistent example of \textbf{Considering Alternatives} was found in the AI interfaces for Richard, all of which mentioned diuretics in some way.
Twelve of 19 participants who saw one of these plan mentions considered starting diuretics, compared to only one of 6 participants who did not see an AI.
Though the results of considering this plan and the perception of its usefulness differed, nine of the twelve participants did ultimately give diuretics.
For example, P13 described how their decision shifted due to seeing the option to give diuretics:
\begin{quote}
    \textit{``I wouldn't necessarily have thought about [diuretics] unless the respiratory status worsened... I tend to err on the side of kind of just like watching and waiting if there's nothing like pressing me to do something... but yeah, he is quite positive and with worsening respiratory status. So I think it would be reasonable to give diuretics and kind of watch closely.''}
\end{quote}

Reasoning cues containing predictions or plan mentions also affected participants' hypotheses and plans more indirectly through \textbf{Inspiration}.
Unlike Considering Alternatives, Inspiration occurred when the participant noticed something surprising in the AI output that led them to form new lines of reasoning that drove their final plan. 
For example, P20 expressed surprise and confusion at a \diffriskyes{} (difference in plan-dependent risk) cue from the Interactive Treatment Risk interface showing that stopping both fluids and vasopressors would have the most beneficial outcome.
However, they soon arrived at a possible explanation: \textit{``Very frequently, the patients go on propofol [for mechanical ventilation]. Propofol’s great and all, but it can cause bradycardia and hypotension... This would be very consistent with propofol if her disease process is not so severe to be causing this.''}
As a result of this hypothesis, they decided to give some fluids (against the AI insight) but then stop treatment if the patient's sepsis seemed to be improving on its own.
Conversely, several AI-recommended plans (\recommendedplan{}) that initially seemed surprising led participants to generate new justifications used to ultimately align with the AI.
Inspiration by a reasoning cue spurred more in-depth thought processes that then proceeded \textit{independently} of the cue.

Information showing a plan-dependent difference in outcomes (\diffriskyes{}) or peer actions (\commonactionno{}) sometimes led participants into \textbf{Second-Guessing} their initial proposals.
Participants were particularly swayed to reject their plans after seeing that very few previous clinicians took an action, as we explicitly observed in four cases.
For example, P7 reacted to a \commonactionno{} reasoning cue as follows: \textit{``My inclination would be to stop IV fluids and give diuretics. [selects on interface] Interesting... pretty much nobody did that.''}
Ultimately, the reasoning cue prompted this participant to consider other ways to mitigate fluid overload without affecting the other aspects of their plan.

\subsubsection{Updating Goals \& Assessment}

Participants most often made their initial assessments of the patient's state automatically as they reviewed the available data.
However, participants sometimes \textit{updated} their assessments through patterns such as \textbf{Resolving Contradictions} when the initial impression revealed seemingly contradictory signs.
For example, the unusual features cue (\unusualfeatures{}) in the Case Features interface flagged Deborah as having a history of valvular (heart) disease, which could point against administering more fluids; however, the vignette mentioned that the patient's most recent echocardiogram showed no current valvular disease.
4/5 participants who saw this cue therefore spent additional time looking for other signals to make sense of the conflicting heart function indicators, ultimately deciding that the prior valvular disease was resolved and the patient could tolerate more fluids: \textit{``I'm getting less of a sense of heart failure''} (P21).
In contrast, participants in other AI conditions were often more cautious in their fluid decisions for Deborah, and none out of the six clinicians who saw no AI for this case decided to give more than 1 liter of fluids.
In this way, \unusualfeatures{} focused participants' attention on particular issues during their assessment and enabled them to make a bolder decision (even though it was opposite what the AI would have suggested).
Notably, we did not observe any influences of the consistent features reasoning cue (\consistentfeatures{}), indicating that clinicians were most focused on the potential anomalies in a case.

Another important and often automatic aspect of assessing the patient was \textbf{Uncertainty Calibration}, or deciding how much to prioritize information-gathering in their next steps for the patient.
A common pattern we saw throughout both studies was that rather than expressing uncertainty about a decision, clinicians often decided to use trial therapies (e.g., a small amount of fluids) to collect additional information for future actions.
As summarized by C6, \textit{``response to treatment is diagnostic information.''}
In Study 2, moderate risk scores (\moderaterisk{}) helped clinicians prioritize information-gathering decisions, such as in Heather's case:
\begin{quote}
    \textit{``[reading AI] `As likely as not to require vasopressors' - I agree. I think she's sort of like declaring herself... I think she's one of those that I would probably, in order to like try and figure out if she's gonna respond to more volume or not... maybe give her like a rapid 250, 500 CC [fluid] bolus and see how she responds.''} (P4)
\end{quote}

\subsubsection{Agreeing, Disagreeing, and In Between}
Participants very often expressed either agreement or disagreement with AI information; however, even these seemingly-simple reactions to the AI had varying effects on reasoning.
The most common form of agreement was \textbf{Confirmation}, where a clinician already had an assessment or plan that the AI aligned with.
Participants' confidence could be increased by seeing a wide variety of reasoning cues, including risk predictions (\lowhighrisk{}), differences in plan-dependent predictions (\diffriskyes{}), peer actions (\commonactionyes{} or \consensusactionyes{}), or plan recommendations (\recommendedplan{}). 
For example, upon seeing that their decision not to give fluids coincided with that of most clinicians, P3 stated that the \commonactionyes{} cue on the Prior Clinician Actions interface \textit{``gives me a little reassurance that what my inclination is... is what's commonly done.''}
Similarly, P5 noted that the prediction that Esther would not require vasopressors (\lowhighrisk{}) \textit{``made me more confident in my plan to try to give her some fluid first before giving her pressors.''}
(In many other cases, participants expressed agreement with a reasoning cue after the fact but denied it having any influence at all on their decision; we do not consider these instances of Confirmation.)

A stronger but more infrequent form of agreement was \textbf{Acceptance}, where a clinician had not made a plan yet and decided to adopt the plan provided by the AI.
For example, all six participants who saw the Treatment Recommendation interface for Richard accepted its recommended plan (\recommendedplan{}) of weaning off the second vasopressor, since this decision was consistent with most participants' assessments across conditions. 
Similarly, one participant adjusted the timing of giving vasopressors to align with the consensus peer action (\consensusactionyes{}): \textit{``Maybe start a touch of [vasopressor] if it doesn't seem like her [blood pressure]'s above 60, but just kinda... giving some time for the antibiotics to work... I feel like I would have probably went sooner to vasopressors in this case than the AI recommended''} (P15).
Acceptance of AI information that did \textit{not} align with participants' initial assessments was much rarer, and only occurred after an Inspiration that led the participant to a new assessment.

Rejection of plans proposed by the AI was also extremely common. 
While rejection often led to dead-ends in participants' thought processes, in some cases they gave a \textbf{Justified Rejection} that ultimately moved their decision process forward.
For instance, many clinicians expressed disagreement with insights for Deborah and Heather that the most favorable decision (\recommendedplan{}) was to not give fluids or vasopressors.
P22 reacted by considering reasons why vasopressors were needed for Heather, leading them to a possible diagnosis that further justified their decision:
\begin{quote}
\textit{``I disagree with the [AI]. This is someone who... especially with that [kidney injury], may be in hepatorenal syndrome needing a higher [blood pressure]. And oftentimes we rely on pressors... I probably would give upfront vasopressors... because she has hepatorenal syndrome with that creatinine, [and] she’s not making urine.''}
\end{quote}
Participants also sometimes rejected information about consensus peer actions (\consensusactionyes{}) by contrasting their preferences with shortcomings they perceived in other clinicians, such as not knowing when to watch and wait: 
\textit{``It is funny to me to see... if somebody felt like they had to do something one way or another, when in reality, like if we probably just hang on for a second, like, things will continue to resolve''} (P2).
In contrast to Second-Guessing, rejection often occurred because participants were unable to rationalize cues related to outcomes (\anyrisk{}, \diffrisk{}) if they were provided.
Instead, they ignored these outcome predictions and provided justifications for their own thought process that helped solidify their reasoning.

Some plans recommended by the AI (\recommendedplan{}) served as a basis for \textbf{Adaptation}, where clinicians adjusted the plan to better align with their practice style.
Adaptation often occurred when an AI plan represented a more aggressive treatment strategy than the clinician would ordinarily take, such as a recommendation to add a secondary vasopressor for Deborah.
Although many participants agreed that Deborah might eventually need an additional vasopressor, they often wanted to adjust the timing of this treatment or make it contingent on the results of another action: \textit{``I would hold off on the additional pressor at this time. I’d wait to see how she responded to the fluids. If she does not respond to fluids... then I think I would add on [the second vasopressor]''} (P19).

\subsubsection{Evaluating and Comparing Plans}
The most in-depth reasoning prompted by our reasoning cues occurred when participants had multiple plausible plans in mind, leading them to use the AI to differentiate between them.
As described in Sec. \ref{sec:study-context} and echoed throughout our observations, many aspects of critical care decisions can have multiple alternatives with no clear right answer across all patients (C3, C4, C5).
Seeing improved outcome predictions for one alternative over another for a patient, therefore, could lead participants to choose the treatment with better outcomes (a pattern we call \textbf{Plan Preference}).
For example, P5 browsed the differences in risk (\diffriskyes{}) for doing nothing and giving diuretics on the Interactive Mortality Risk interface for Richard, and made their decision based on the predicted outcomes:
\textit{``It seems like the [AI] is suggesting based on data from 100 patients, I guess they say that giving diuretics is the way to go.''}
Clinicians also frequently chose between treatments if one was shown to be more common than the other.
In Esther's case, many participants weighed giving either vasopressors or fluids; all of those who saw a \consensusactionyes{} (consensus action) cue for this patient ended up giving fluids because it showed most clinicians not giving vasopressors.

On the other hand, plan-dependent predictions and peer actions with \textit{no} meaningful differences between the plausible options sometimes led to a pattern called \textbf{Acknowledging Equipoise} (referring to the clinical concept of no treatment option being obviously more effective than another). 
This reasoning pattern was similar to Confirmation in its effects on participants' confidence, but it did not require the reasoning cue to explicitly align with their assessment.
Some participants found it reassuring that the vasopressor treatment options for Deborah showed no significant impact on mortality (\diffriskno{}), allowing them to take the decision most consistent with their usual practice: \textit{``It sounds like you could have done anything here... and it wouldn't have changed much''} (P3).
In general participants rarely considered multiple viable treatment paths, perhaps because they often managed any clinical uncertainties by opting for information-gathering actions (C1, C6).
With these reasoning cues, however, participants were more likely to choose a definitive path while acknowledging the presence of others: 
\begin{quote}
    \textit{``I think what [the lack of consensus actions \consensusactionno] really shows is that there's a variety of strategies here, that there's probably not one right answer. You've just got to, like, choose one and go with it.''} (P12)
\end{quote}

\subsection{Perceptions of Reasoning Cues}
\label{sec:results-perceptions}

Thematic analysis of our contextual inquiry observations and think-aloud study feedback revealed insights for \ref{rq:perception} that inform the design of reasoning cues both for sepsis and human-AI decision-making more broadly: supporting the \textit{right} reasoning tasks, being \textit{compatible} with the clinician's current information needs, and providing cues that are \textit{complementary} to guidelines and existing knowledge.

\subsubsection{Supporting the Right Reasoning Tasks: Focusing on Variability and Discretion}
\label{sec:perceptions-right-tasks}

Participants throughout both studies distinguished between critical care decisions that are largely protocolized and those that depend heavily on clinician judgment. 
On the protocolized side, several described having a low threshold for initiating fluid resuscitation, vasopressors, and antibiotics upon the first suspicion of sepsis: \textit{``Delay in identifying sepsis and appropriate therapies is probably much more detrimental than being heavy handed up front''} (C5).
Intelligent reasoning cues were perceived as unlikely to be able to shift clinicians' management at these decision points because decisions were dictated by guidelines and hospital-level doctrine.

On the other hand, decisions such as how much fluid to give a patient could be highly discretionary and variable among providers. 
Several expert clinicians in Study 1 noted that they had a particular approach to fluid decisions but that others might differ (C1, C2, C3, C5, C7).
This variability was exacerbated in more specialized ICUs where patients are more complex: \textit{“There’s actually no evidence for a lot of what we do… This unit’s an evidence-free zone”} (C4). 
Participants described using a range of techniques to make fluid responsiveness decisions more objective, such as point-of-care ultrasounds, passive leg raise tests, and physical examination. 
However, the limitations of these measures were well recognized, especially because the data they give could be up to interpretation. 
In a setting where even so-called objective measures could lead individuals to develop different interpretations and plans, reasoning cues that predicted the patient's current fluid responsiveness were suggested as a potential tool to structure decisions (C2). 


Another notable contrast between protocolized and discretionary decisions was in early versus late sepsis management.
For example, patients in the ICU often require multiple supportive therapies such as invasive ventilation, vasopressors, and kidney support, but it is not clear how and when these therapies should be \textit{removed} as the patient's condition improves.
In many cases we observed clinicians opting to remove treatments one at a time to ensure that the patient remained stable (C2, C3, C4).
However, some participants found it helpful when a reasoning cue such as \diffriskyes{} presented a difference in plan-dependent risk that might push them to wean multiple therapies simultaneously: \textit{``I feel like diuresing someone on pressors always feels like a little bit of a bold thing... It helped to have [the AI] say that that... was more likely than not to lead to survival''} (P5).
Later stages of a patient's course could also be challenging because even though more data was available, the lack of improvement in a patient's condition was an indication of more elusive underlying causes (C1, C5, C6, C7). 
For example, one physician described that patients who were several days into an ICU stay and still on vasopressors triggered greater uncertainty:\textit{``Have we considered all sources of infection? Other causes of shock?''} (C6).

At challenging decision points such as these, descriptive reasoning cues such as \unusualfeatures{} (unusual features) had potential to support patient re-assessments.
Participants in Study 1 suggested they would be interested in AI that could gently disrupt faulty reasoning patterns: \textit{''You’d like something to short-circuit [your heuristics] and say, ‘Hey, take a step back and pause''} (C6).
In Study 2, however, participants noted that such tools would only be helpful \textit{after} they had formed their own impression: \textit{''I still want to look through first myself and not rely on someone to tell me [the patient's condition]''} (P23). 
Participants were particularly interested in seeing \unusualfeatures{} at times when a re-assessment was needed, such as the emergence of new data trends or a lack of improvement.
For example, in reaction to \unusualfeatures{} flagging Heather's heart rate, P3 explained:
\begin{quote}
\textit{``I would treat [Heather] the same way as I would treat a person with a normal heart rate. And only if after that first-pass assessment, things aren't going well or things are different, I would as a second pass, think a little bit harder about her heart rate.''}
\end{quote}

\subsubsection{Compatibility with Information Needs: Adapting to Case Contexts}
\label{sec:perceptions-compatibility}
Clinicians in both studies emphasized that intelligent reasoning cues must be \textit{compatible} with the ways they already reason through decisions.
They often expressed that the same reasoning cues could vary significantly in usefulness in different specialties and stages of reasoning.
For instance, one participant welcomed the mortality predictions shown by \anyrisk{} and \diffrisk{}, saying, \textit{``In this clinical context, mortality is the appropriate outcome... If they don't live, none of those other things matter''} (P8). 
But another pushed back: \textit{``I don’t necessarily think about these patients as a mortality risk all the time... I think more like, how are we going to get them off pressors?''} (P1). 
Participants who found mortality less useful suggested other targets that could improve the \anyrisk{} and \diffrisk{} cues, such as the risk of intubation or the likelihood of being able to stop vasopressors.
They also pointed to circumstances in a patient's care, like their initial emergency department presentation or goals of care conversations with family, in which mortality \textit{could} be more helpful.
These divergent perspectives highlight how the usefulness of reasoning cues was not static, but depended on whether the output was compatible with the goals the clinician had in mind. 

Compatibility also included whether the mentioned plans (\planmention{}) were aligned with the options they were considering. 
Several participants expressed frustration when our AI interfaces suggested irrelevant or unrealistic options: \textit{``I want to see it predict what I'm gonna do, and based on that, delete the irrelevant options... For example, if reasonable people would do A and B, and... will not choose C and D, why do I even have to look at C and D?''} (P25).
Conversely, participants often found it unhelpful when the Interactive interfaces had insufficient data (\commonactionno{}) for treatment options they thought were promising: \textit{``I felt like [this interface] had the potential to be the most useful. But it just never really provided an answer''} (P23).
From these participants' perspective, the ideal role of the AI was in helping them evaluate the plans they were already considering, and only presenting the possibility of new plans when they were deemed \textit{viable} based on the patient's immediate needs.
As P8 put it, \textit{``I would be unenthusiastic to [follow this \planmention{} cue] because I think that if we were to do so, she would drop into a blood pressure range that's incompatible with life.''}

\subsubsection{Complementing Existing Knowledge and Guidelines}
\label{sec:perceptions-complementarity}

Many participants were excited for the potential of AI to provide new information to support their decisions, but having new information was not \textit{always} desirable.
As described in Sec. \ref{sec:perceptions-right-tasks}, decision aspects that are already highly driven by guidelines and protocols are unlikely to be influenced by intelligent reasoning cues.
In fact, for guideline-driven decisions, participants expected the AI not to even attempt to offer novel information or influence reasoning; instead, they wanted it to confirm and validate their actions.
When participants' decisions matched both guideline-based best practice and the AI information, they noted that the AI \textit{“helps you have an extra level of confidence when everything matches what you have”} (P2). 
This sentiment was echoed by clinicians in Study 1 who suggested a decision tool that presents and enforces protocolized treatment plans \textit{``for people who don’t have the guidelines all memorized''} (C2). 
Overall, clinicians only valued AI in straightforward decisions if it reinforced the guideline-based decision structures they already trust. 
Of course, all the cases in Study 2 were chosen because of their complexity, underscoring that a clinician has to assess the patient as being complex in order to be open to reasoning support.


Although clinicians saw the most potential for complementarity when cases were more nuanced or atypical, their assumptions about AI systems often led them to question whether our reasoning cues could truly account for the full complexity of these patients. 
Experts in Study 1 emphasized the importance of balancing treatment priorities across multiple organ systems, integrating specialist perspectives, and aiming for the best overall outcomes. 
As mentioned in Sec. \ref{sec:study-context}, these priorities often extend beyond the scope of standardized sepsis guidelines (and the purview of our AI system). 
For example, one clinician from an ICU focused on neurological conditions described how preserving brain function can sometimes take precedence over targets set by sepsis protocols when determining blood pressure goals (C4).
In such cases during Study 2, doubts about AI's ability to factor in all relevant clinical factors shaped whether and how participants acted on its recommendations.
For instance, P24 assessed Heather's case as being worsened by hepatorenal syndrome (HRS), and they wished the AI could recognize that risk and bring in more specific information: \textit{``I don’t think the data is there yet for AI to make treatment decisions [for patients with both sepsis and HRS].''}

In complex cases where existing guidelines might not be sufficient, participants most strongly preferred reasoning cues that provided what they called \textit{“true data”} (P14), defined as concrete, context-specific evidence based on intrinsically interpretable statistics on patients. 
Rather than broad recommendations, they sought differences between plans (\diffrisk{}, \commonaction{}) based on identifiable studies, relevant patient populations, or aggregated patterns from similar cases. 
Some likened these reasoning cues to differing preferences between care team members, noting that in such cases they would more likely to be comfortable with the action favored by the AI.
Others envisioned AI coalescing real-world practice patterns in the absence of trials, with P24 noting, \textit{“It’s useful to me because there’s no study on that.”} 
Finally, reasoning cues that helped evaluate plans were seen as promising because they could inform deviation from protocols that participants knew were somewhat arbitrary:
\begin{quote}
\textit{``Most of the things that we work [on] in critical care... they're evidence-based, but they cannot predict the future. [A treatment] can go bad or it can be good, even though this is standard of care. So if it can remove the subjectivity from that decision-making... that'll be excellent.''} (P25)
\end{quote}
Similar to the results above, seeing value in this plan evaluation process depended on clinicians arriving at multiple plausible options from earlier stages of reasoning. 
But when they were, they could leverage cues based on true data to fill the gaps left by guidelines, weigh trade-offs and make more confident decisions when the standard playbook ran out.

\section{Discussion and Limitations}

In this work we developed and presented intelligent reasoning cues, a framework for how AI-derived information integrates into human reasoning on complex decisions.
Although many studies have developed new designs for AI-based decision support~\cite{reicherts_ai_2025,zhang_resilience_2023,Henry2022} or attempted to understand what makes these systems effective or not~\cite{medlock_modeling_2016,kawamoto_improving_2005,kawakami_why_2022,sivaraman_ignore_2023,Beede2020}, our work is among the first to establish a direct connection between the specific informational elements produced by AI and the mechanisms by which they influence reasoning.
We did so by decomposing an AI interface into multiple reasoning cues and analyzing each cue's influence on decision-makers' hypotheses, uncertainties, plans, and perceptions.
Our case study of sepsis treatment decision-making illustrates that we can trace reasoning cues to specific patterns of influence as well as perceptions about the value of different AI information types. 
These results point to several ways that the intelligent reasoning cues framework can provide a basis for future AI-based decision support design.

\paragraph{Improving Decision Support Designs in Sepsis}
Our work indicates several future directions for the sepsis treatment context that may generalize to other contexts as well.
Most notably, critical care clinicians experienced friction when the AI intervened in relatively straightforward and standard decisions (such as giving fluids and vasopressors in response to low blood pressure) compared to tasks with greater variability and discretion (such as deciding how much fluid to give).
In several instances in Study 2, however, cases selected for their complexity were perceived as straightforward and standard by many participants.
This suggests that intelligent reasoning cues for sepsis treatment would be most effective when decision-makers already perceive a range of viable options.
A promising strategy to encourage this frame of mind was to show peer actions (particularly the presence of a consensus action \consensusactionyes{}), which led many to Second-Guess their initial choice.
Once open to alternative decisions, flagging unusual features (\unusualfeatures{}) helped participants update their initial assessments, while plan-dependent risk differences (\diffrisk{}) and peer action frequencies (\commonaction{}) allowed them to weigh multiple options.
Each of these patterns of influence could play a role in improving overall decision quality, and could therefore be promising interventions to test in a quantitative evaluation.

Our findings confirm and expand on prior work in sepsis treatment and other high-stakes domains.
For example, our Adaptation influence pattern aligns with \citeauthor{sivaraman_ignore_2023}'s ``Negotiate'' behavior group, in which clinicians chose some aspects of the AI's recommendation to rely on but rejected others or navigated an intermediate path ~\cite{sivaraman_ignore_2023}.
However, their results grouped several different behaviors related to negotiation at the participant level, while our analysis method allowed us to identify behaviors at the level of individual hypotheses and plans.
The influence patterns we identify are also similar to \citeauthor{zhang_resilience_2023}'s model of pilots' action needs and means of supporting them~\cite{zhang_resilience_2023}, although their findings are based on perceptions rather than behavioral observations.

Our results also complicate prior findings on the types of cases where expert decision-makers most value AI support.
Participants in our study expected systems to encode and enforce guidelines when decisions are considered straightforward, which could serve to help justify decisions and comply with administrative constraints~\cite{kawakami_improving_2022}.
Yet in decisions with greater discretion, where decision-makers have historically preferred to rely on their own judgment~\cite{chen_understanding_2023,kawakami_why_2022}, clinicians in our study described being much more interested in using an intelligent reasoning cue if it provided \textit{accurate retrospective evidence} going beyond clinical literature.
Using an intrinsically-interpretable nearest neighbor approach on an underlying model, as we did to generate our reasoning cues, could be a promising way to create such evidence and tailor it to an individual decision. 

\paragraph{Using Reasoning Influence Patterns to Guide Decision Support Design.}
Our study results suggest that users engage with intelligent reasoning cues through specific patterns of influence, and that the value they derive from a cue depends on how the cue is formulated and the influences it exerts.
Based on these findings, we propose a possible workflow for designing AI interfaces to optimize the reasoning cues they deliver:

\begin{enumerate}
    \item \textbf{Identify the right tasks to support.} HCI research has provided several methods to elicit promising ideas for AI support, such as collaborative sketching~\cite{yildirim_sketching_2024} and co-design~\cite{panigutti_co-design_2023}. For example, Study 1 suggested that the most promising tasks to support in our context were those that are \textit{highly variable among providers} and \textit{poorly supported by guidelines} (Sec. \ref{sec:perceptions-right-tasks}), an observation that may hold true in other domains as well. 
    \item \textbf{Identify desirable patterns of influence.} At this stage, an approximate model is needed to understand how decision-makers might reason about the tasks of interest. Fig. \ref{fig:decision-workflow} shows the stages of reasoning observed in our results and influence patterns that could occur at each stage; different stages may be observed in contexts beyond clinical treatment decisions. Using this type of reasoning model, we can choose to design for influence patterns that would overcome observed flaws in the reasoning process. For instance, if decision-makers often fail to make sense of conflicting information, we might choose to build a tool for Resolving Contradictions.
    \item \textbf{Select reasoning cues that promote the desired pattern(s) of influence during the decision-making process.} Given a set of influences, the findings of our study (Fig. \ref{fig:decision-workflow}) point to specific reasoning cues that are likely to promote those influences. This is the key step enabled by the intelligent reasoning cues framework.
    \item \textbf{Implement and refine reasoning cue designs with stakeholder input.} At this point, we can apply a variety of existing technical and human-centered methods~\cite{savage_user_1996} to develop and refine an interface containing the reasoning cues. The criteria described in Sec. \ref{sec:results-perceptions} can help structure iterative feedback from stakeholders and potential users: Are the reasoning cues supporting the right tasks? Are they compatible with the user's context and complementary to their existing knowledge?
\end{enumerate}

\paragraph{Generalizing Findings to Other Human-AI Decision-Making Contexts.}
The specific reasoning cues we designed and the influence patterns we observed are merely a starting point for future research, given that they are based in a single decision context. 
Although our study covered a broader range of reasoning cues than any single prior experiment to our knowledge, we only tested those that could support clinical treatment decisions.
Further reasoning studies with tasks such as prioritizing patient cases~\cite{Henry2022,Sendak2020} and making value-laden judgments~\cite{kawakami_why_2022} could expand the toolbox of reasoning cues and influence patterns available to interaction designers.
Furthermore, our think-aloud study focused on fellows, who are trained to make treatment decisions but may use different reasoning patterns than more experienced providers.
While we attempted to mitigate this possible source of bias by capturing a wider range of real-world reasoning patterns in Study 1, further research is needed to understand whether the same cues might induce different patterns of influence for people with different levels of expertise.

An important area in which to explore our framework's generalizability is the space of new reasoning cues enabled by GenAI systems.
While the unstructured nature of GenAI-based decision support systems allows them to adapt more seamlessly to a decision-maker's current context and needs~\cite{reicherts_ai_2025,ma_towards_2025}, our findings suggest that it may also be important to \textit{structure} their output by prompting them to provide reasoning cues grounded in ``true data.''
We also observe that decision-makers can be positively influenced by intelligent reasoning cues that they did not explicitly request, such as \unusualfeatures{} \unusualfeatures*{} and \commonaction{} \commonaction*{}.
Whereas current GenAI systems might require the user to articulate their needs, systems based on reasoning cues could have stronger priors based on desired patterns of influence.
Better strategies to combine the priorities of the decision-maker with those of the system designer could be key to driving improvement of GenAI-assisted decisions~\cite{goh_large_2024}.

\paragraph{Making Sense of Mixed Human-AI Decision-Making Results.}
Intelligent reasoning cues can also help us re-contextualize the findings of past studies of human-AI decision-making to inform future design efforts.
For instance, let us consider two studies comparing the effects of feature-based and example-based AI explanations: a mixed-methods study on two general-knowledge tasks by \citet{chen_understanding_2023}, and a quantitative study of chest X-ray diagnosis by \citet{prinster_care_2024}.
At surface level, these experiments yielded contradictory results: \citeauthor{chen_understanding_2023}'s study suggests that example-based explanations lead to better decision-making performance while \citeauthor{prinster_care_2024}'s results suggest that feature-based explanations are superior.
(Both studies consider varying levels of expertise with respect to the task, so expertise cannot be the sole cause of the difference.)

We can begin to resolve this contradiction by examining the reasoning cues embedded in each study's AI interfaces.
In the former, the feature-based explanation gave equal visual weight to all features, whereas in the latter the feature explanation directly highlighted one region of the chest X-ray to direct the user's attention (similar to our unusual features cue, \unusualfeatures{}).
For the example-based explanation, the former's interface crucially included multiple examples along with the AI's correctness, while the latter showed only a single similar instance.
As \citeauthor{chen_understanding_2023} themselves point out, the added information provided by their example-based explanation likely helped users form intuitions about the task that improved their performance.
Meanwhile, the same could be said of the feature explanation in \citeauthor{prinster_care_2024}'s study, since it directed non-experts' attention to relevant parts of the image.
Thus, although both of these studies advance our understanding of human-AI interaction, neither can conclude an inherent benefit of one explanation type over another because their results are confounded by the presence of additional reasoning cues.
This example also underscores how explainable AI can have strong potential usefulness and impact, but its success depends on the specific reasoning cues that are used. 
Our framework can help researchers better pinpoint what cues would best support reasoning in their domains, allowing them to make better use of the wealth of existing research on human-AI decision-making.

\paragraph{Measuring the Impact of Reasoning Cues.}
As AI-driven decision support systems are being developed and deployed in high-stakes settings, it is important to determine which system designs quantitatively improve decision-makers' accuracy, efficiency, and satisfaction in their work. 
In prior human-AI decision-making experiments, however, it has often been difficult to untangle the effects of the desired interventions (such as explanations) from those of other pieces of information they may carry with them~\cite{Wang2019,papenmeier_its_2022}. 
Given our focus on collecting rich information on clinical reasoning and perceptions, we did not attempt to quantitatively evaluate reasoning cues in this work.
 Our findings point to promising cues and influence patterns that should be tested in future studies.
To mitigate confounding across different AI designs, such studies could test ablations of the same AI system with reasoning cues successively removed to see which cues provide the most benefit.

\section{Conclusion}
The growing range of capabilities that AI systems can perform accurately has great potential benefits for high-stakes decisions.
However, general principles of how to best design AI for decision-making have largely remained elusive~\cite{vaccaro_when_2024,gomez_human-ai_2025}.
Our work attempts to bring clarity to this issue by building a rich, qualitative understanding of how AI information influences clinicians' reasoning, which may translate to similar processes performed by other kinds of experts.
Furthermore, we demonstrate the utility of considering AI interfaces in terms of intelligent reasoning cues, which can be traced to specific patterns of influence within a decision-making process.
Through the lens of intelligent reasoning cues, insights from human-AI decision-making research could become a more practical guide to designing AI systems for important and complex decisions.

\begin{acks}
We are grateful to the many critical care physicians, nurses, and other providers who allowed us to observe their practice and participated in our interviews and think-aloud sessions.
We also thank Katelyn Morrison, Arpit Mathur, Dominik Moritz, Peter Nauka, Sivaraman Sivaswami, Haiyi Zhu, Mayank Goel, and Suchi Saria for their critical feedback on the framework and study design.
This work was supported by a research grant from the United States National Institutes of Health (R35HL144804) and a National Science Foundation Graduate Research Fellowship (DGE2140739).
\end{acks}

\bibliographystyle{ACM-Reference-Format}
\bibliography{main}

\appendix

\section{Additional Study 2 Details}
\label{app:study-2-details}

\subsection{Study Interface}
\label{app:study-interface}

\begin{figure*}
    \centering
    \includegraphics[width=\linewidth]{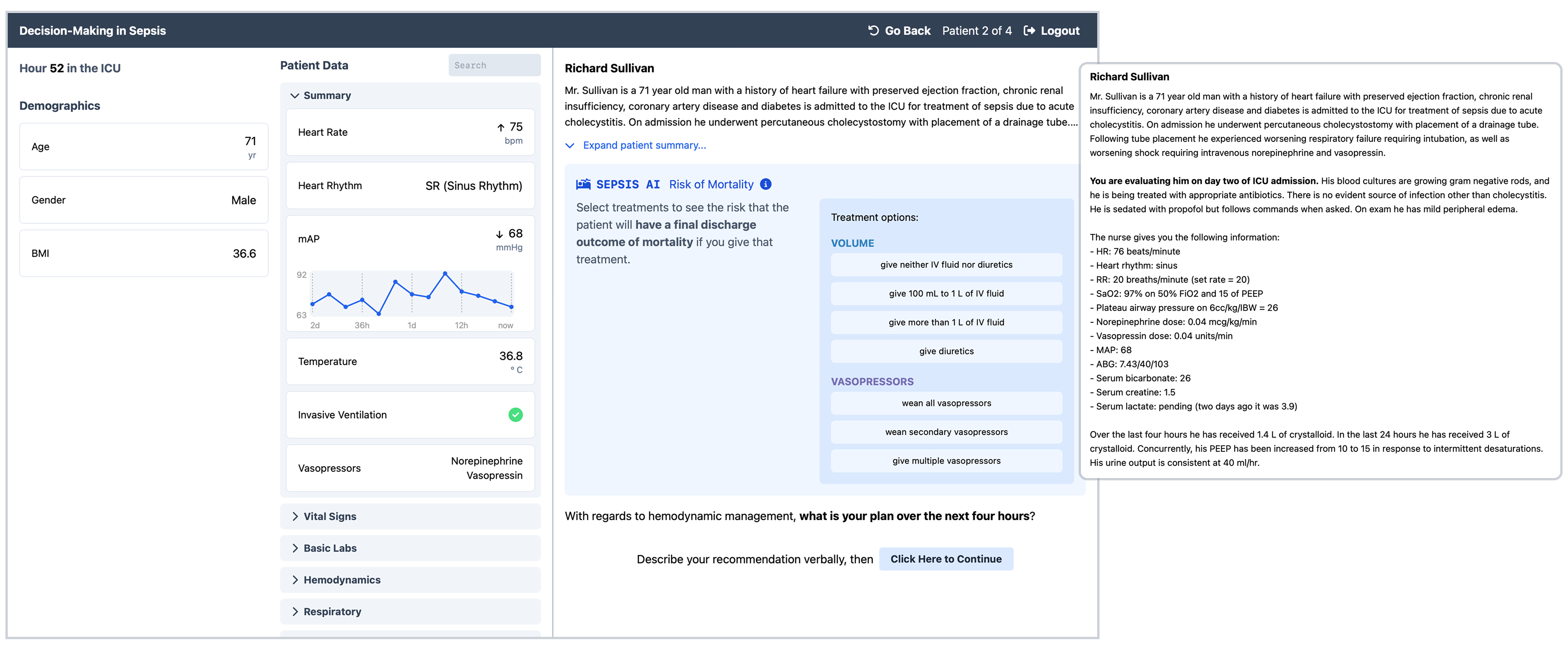}
    \caption{Example of the interface and vignettes shown to participants in Study 2. The blue box labeled Sepsis AI would contain the AI interface and reasoning cues they were randomized to receive for the given patient.}
    \label{fig:study-interface}
\end{figure*}

Shown in Fig. \ref{fig:study-interface}, the interface that was shown to participants in Study 2 consisted of three parts: structured data from the patient's health record, the unstructured case vignette, and the AI interface (if applicable).
Each patient's information was drawn from MIMIC-IV~\cite{johnson2020mimic}, which contains ICU trajectories for real patients seen between 2008 and 2019.
For each patient, the structured data (left half) included the patient's demographics and an expandable list of 56 data indicators, including their vitals, lab values, hemodynamic and respiratory status, and prior treatments. 
Each indicator showed the current value as well as the history of the value over the course of the patient's ICU stay.
Similar to existing EHR systems, an arrow indicated each value's recent trend and the values were colored in red if they fell outside normal ranges.
We iteratively developed these categorizations and presentations in collaboration with a clinical expert; most participants were able to review all indicators for every case within the allotted time.

The text vignettes were roughly 200-300 words each, and conveyed patients' status in a style that combined elements of a medical exam question with a standard patient presentation.
To simulate the information that would be available to the clinician at the bedside, we used both the structured EHR data available in MIMIC-IV as well as their discharge notes, which contain contextual information known to be essential in making realistic decisions ~\cite{munroe_understanding_2024,sivaraman_ignore_2023}.

Participants were introduced to the interface and the information it contained prior to making their first decision.
They were not required to enter anything into the interface; rather, the system displayed the case information and debriefing questions for participants to answer verbally.
This helped us maintain consistency in question presentation while minimizing participant burden and allowing them to explain their answers freely.

\subsection{Analysis Methods}
\label{app:analysis-methods}

Below are examples from the thematic analysis of the think-aloud sessions used to answer RQ1. To illustrate our methodology, these examples follow the analysis process from initial open coding of one think-aloud session (P5), through further affinity mapping and synthesis into themes.

\begin{figure*}
    \centering
    \includegraphics[width=0.8\linewidth]{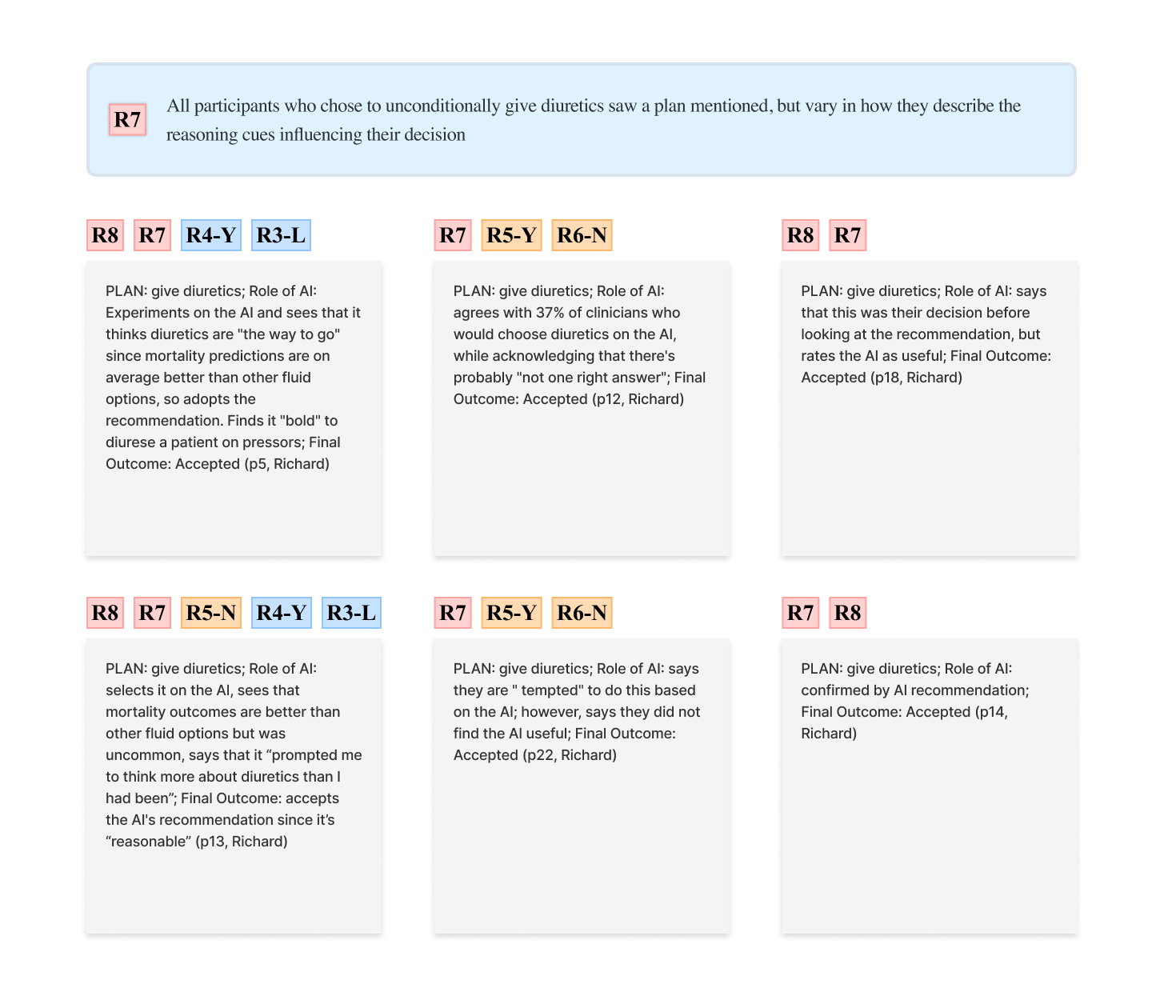}
    \caption{A theme generated from the first round of affinity mapping of codes for each patient case. This example is from Richard's patient case.}
    \label{fig:affinity-mapping-patients}
    \includegraphics[width=0.8\linewidth]{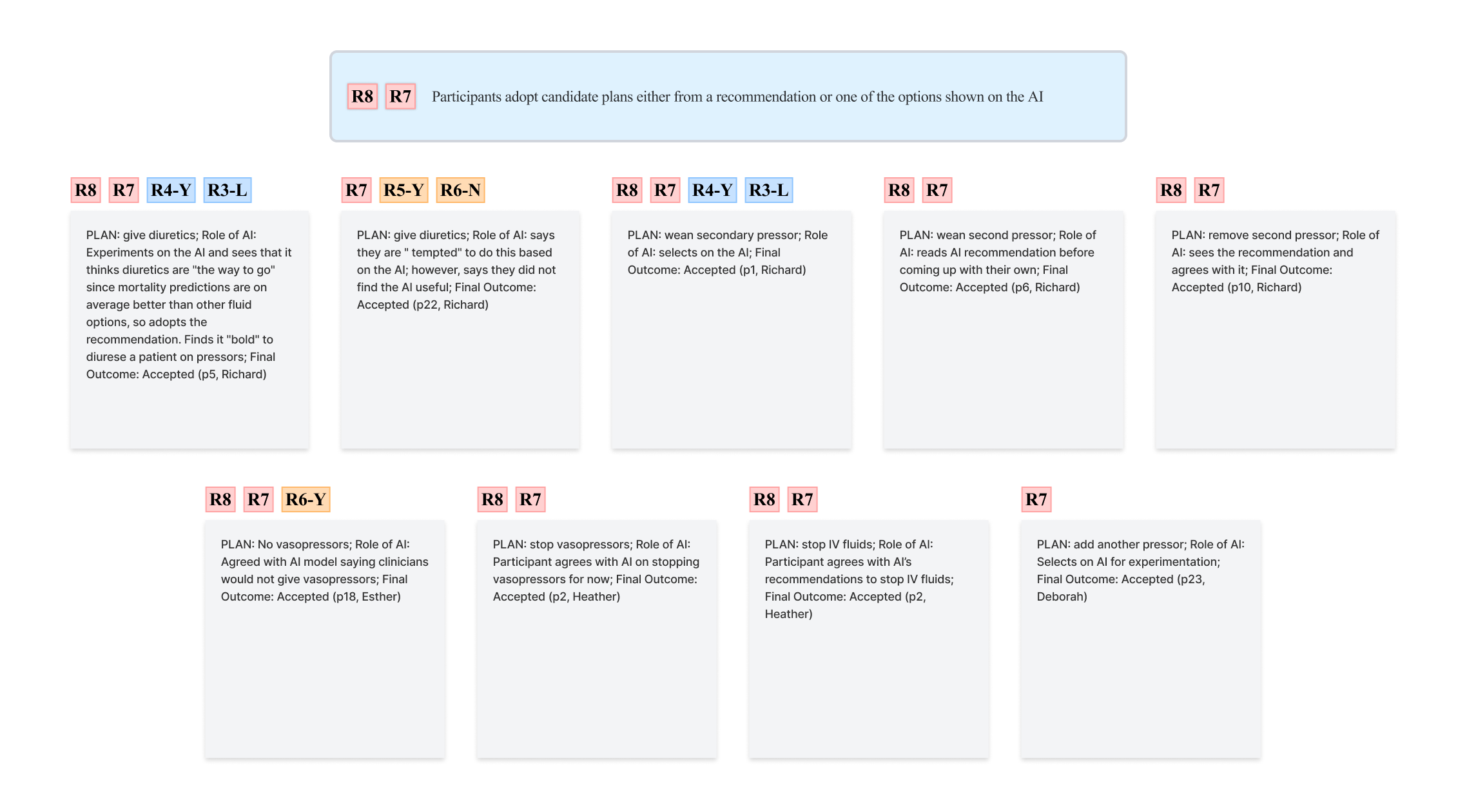}
    \caption{A theme generated from the second round of affinity mapping of codes with the role of AI across all patient cases.}
    \label{fig:affinity-mapping-role-of-ai}
\end{figure*}

\begin{table*}[]
\small
    \centering
    \begin{tabular}{p{2.5cm}|p{2cm}|p{2cm}|p{3cm}|p{3cm}}
    \toprule
\textbf{Hypothesis} & \textbf{Plan} & \textbf{Reasoning Cues} & \textbf{Role of AI} & \textbf{Final Outcome} \\
\toprule

Heart failure and renal insufficiency means patient will be harder to give fluids to & & & & Led to plan of not giving fluid \\
\midrule

 & Don't give fluid & & Later selects as initial plan on AI,  sees that there are not enough patients unless some fluid or diuretics is selected & Accepted \\
\midrule

 & Stop second vasopressor & & Later selects as initial plan on AI, sees that there are not enough patients unless some fluid or diuretics is selected & Rejected - although initially their plan, they reject it after accepting another part of the AI's insights \\
\midrule

 & Continue two vasopressors & \planmention{} \consensusactionno{} \newline \diffriskno{} \moderaterisk{} & Experiments on the AI, sees that  there are not enough patients to generate a prediction unless some fluid or diuretics is selected & Accepted - although stopping second vasopressor was initially their plan, they decide to keep both vasopressors while giving diuretics since they're "not necessarily in a hurry" to get off the vasopressor \\
\midrule


 & Give diuretics & \recommendedplan{} \planmention{} \newline \diffriskyes{} \lowhighrisk{} & Experiments on the AI and sees that it thinks diuretics are "the way to go" since mortality predictions are on average better than other fluid options, so adopts the recommendation. Finds it "bold" to diurese a patient on pressors & Accepted - later states that they find it "bold" to diurese a patient on vasopressors \\

\bottomrule
\end{tabular}
    \caption{Example of how think-aloud transcripts were open coded, from P5 evaluating Richard's case. Note that the first three rows are not associated with any reasoning cues because the participant had not evidently looked at the AI interface yet.}
    \label{tab:interview-coding}
\end{table*}

\subsubsection{Open Coding}
\label{app:analysis-coding}

Shown in Table \ref{tab:interview-coding} is an example of how interview transcripts from the Think-Aloud protocol were coded to capture clinician reasoning processes. 
To assess AI influence on clinician decision-making, we examined screen recordings to annotate which \textit{reasoning cue} participants saw, and described the \textit{role of the AI} if it clearly influenced their thought. 
We then coded when clinicians made a \textit{hypothesis} about the patient's status, expressed an \textit{uncertainty} about the patient's state (omitted in this figure as P5 did not express any uncertainties), or stated an actionable \textit{plan} to treat the patient in terms of volume (IV fluids or diuretics) or vasopressors. 
We also tracked the \textit{final outcome} of whether a hypothesis or plan was ultimately accepted and what role it played in the participant's final decision. 

\subsubsection{Affinity Mapping}
\label{app:affinity-mapping}

Analysis of the codes was performed through two rounds of affinity mapping. In the first round we conducted affinity mapping within each \textit{patient case}. Themes were identified by looking for patterns associated with the types of reasoning cues identified in the codes, as well as in lines of thinking specific to each patient case as hypotheses, uncertainties, and plans were negotiated. This allowed us to not only analyze the explicitly self-reported impact of reasoning cues, but also identify the implicit and potentially sub-conscious influence of reasoning cues on thought processes. For example, Fig. \ref{fig:affinity-mapping-patients} shows a theme produced from affinity mapping codes in Richard's case, where we found a pattern in giving diuretics and seeing diuretics mentioned (\planmention{}), despite participants' differing self-reports on how the AI influenced their decisions. This analysis enabled us to mechanistically understand how AI affected decisions in specific patient cases.

In the second analysis, we conducted affinity mapping on all codes discussing the \textit{role of AI} in across all patient cases. We identified themes in reasoning influences by identifying patterns in reasoning cues and the final outcomes of plans. Fig. \ref{fig:affinity-mapping-role-of-ai} shows a theme generated from codes from all four patient cases where participants adopt plans mentioned on the AI interface, especially recommended plans (\recommendedplan{}). This allowed us to capture the broader patterns of influence of reasoning cues on clinician reasoning across all patient cases.



\subsection{Think-Aloud Protocol}
\label{app:study-protocol}

Below are the text and questions used for our semi-structured think-aloud procedure.

\subsubsection{Introduction}

We’re interested in understanding how physicians reason about treating patients with sepsis, especially using some AI systems. It should take about half an hour to 40 minutes, and you will be compensated \$75. Does that sound good so far?

To help us better analyze the study, this session will be recorded. Are you okay with that?

\subsubsection{Pre-Survey}

First please go to the following link \textit{[link to screening and pre-survey form]} and share your screen.

This first page shows you more detail about the study. Feel free to read over this, and answer the questions at the bottom if you consent. 
Basically, you’ll imagine you are working an ICU shift, and you’ll review information and make recommendations for four different patients. 
After that I will ask you a few questions about the experience, to find additional ways that we might assist with your decision-making. 
We aren’t using any personal identifiers about you in our data analysis, and you can always stop participating at any time.

\textit{[Demographics form]} Before we start looking at patients, we’d like to collect some preliminary information about you to help with our analyses.

\textit{[Tutorial screenshot]} You’re going to be using a page that looks like this to make your decisions. 
We’ll be looking at real patients who are obviously anonymized, but the information you’ll see about them in this tool is real. 
To give a quick overview of the information you’ll have: \textit{[brief description of interface]}

To put you in the mindset of the task, imagine you’re pre-rounding and reviewing these patients and you want to put together a recommendation. 
So you can tell me anything that you would do verbally, and you won’t need to enter anything into the tool.

\subsubsection{Decision-Making Tasks}

Now please click the link in the end of the survey form to go to the decision-making tool.

As we mentioned before, we’re interested in seeing how physicians make decisions for patients with sepsis. 
So your task will be to recommend a treatment plan for each of the four patients, typically in terms of IV fluids, vasopressors, and diuretics. This plan will be carried out over the next four hours. 
As if I were shadowing you, you’ll talk me through how you interpret the information, reason about it, and come up with your recommendation.

During some of the cases, you may see a box labeled Sepsis AI. 
These insights come from an AI system that’s trained on a database of over 14,000 patients to identify similar patients to the one you’re treating. 
It then gives you information about what happened to those patients to support your decision. 
This model has been validated by expert clinicians, and the information it provides is generally accurate, though of course it can make mistakes. 
Do you have any questions about the AI insights before we look at them?

Okay, we’re ready to start looking at patients! 
Remember: I’m shadowing you, so be sure to think out loud and explain to me how you are reasoning about the information you have.

\textit{[Participant reads vignette and makes decision. Experimenters assist in navigating the site where necessary.]}

\textit{[If participant isn’t talking for a while]} What are you thinking right now?

\textit{[If participant wants to do a “collect more information” action]} If we can’t collect that information before rounds, what would your recommendation be given the information we have?

\textit{[After making fluid and pressor decisions]} Now you’ll see a few questions about the decision you just made – go ahead and answer those out loud to me.

\begin{itemize}
    \item How mentally demanding was this case?
Answer on a scale from 1-10, with 1 being not at all demanding and 10 being extremely demanding.
    \item How confident are you in your treatment decision?
Answer on a scale from 1-10, with 1 being not at all confident and 10 being extremely confident.
\item \textit{[If AI was shown]} How useful was the information provided by Sepsis AI for this case?
Answer on a scale from 1-10, with 1 being not at all useful and 10 being extremely useful.
\end{itemize}

\textit{[Continue through the remaining 3 vignettes. Make notes of any interesting behaviors to follow up on during the debrief.]}

\subsubsection{Debrief and Discussion}

That’s all of the patients! I just have a few questions for you to reflect on how you made decisions on those four cases. 

\textit{[First, follow up on any interesting observations that emerged during the study. Then, ask any of the below questions that are relevant given time.]}

\begin{itemize}
    \item You rated [X] AI interface the most helpful. Would you still say that and why?
    \item \textit{[If they saw an AI with multiple treatment options]} Thinking back to the [AI with multiple options], would you expect to see more or less treatment options for patients like these? Did that change your decisions or how confident you felt?
    \item \textit{[If they didn’t see an AI with multiple treatment options]} Would it change your decisions or your confidence if you had more than one treatment option in the AI insights?
    \item If the AI recommendations were more discordant with your assessment, how might that affect your thinking?
    \item How would you compare the decision-making process using each of the AI insights that you saw?
    \item Are there other types of insights that you would find helpful when reasoning about sepsis patients?
    \item How would you compare the decisions you made in this study to the decisions you make on an everyday basis? Were these cases easier or harder than the cases you typically see?
\end{itemize}

That’s all the questions we have for today. Before we close, do you have any other comments or suggestions that you want to give on the interface, the AI, or the decision-making process?

\end{document}